\documentclass[11pt]{amsart}
\usepackage{amssymb,amsthm}

\newcommand{\tr}{{\text{\rm tr}}}
\renewcommand{\d}{\partial}
\def\R{{\bf R}}
\def\C{{\bf C}}

\newcommand{\la}{\langle}               
\newcommand{\ra}{\rangle}               
\newcommand{\eps}{\epsilon}
\newcommand{\half}{\frac{1}{2}}
\newcommand{\ordo}{o}
\newcommand{\ame}{{}^{(4)}g}            
\newcommand{\aR}{{}^{(4)}R}             
\newcommand{\ok}{{}^0k}                 
\newcommand{\og}{{}^0g}                 
\newcommand{\ophi}{{}^0\phi}            
\newcommand{\orho}{{}^0\rho}            
\newcommand{\oj}{{}^0j}                 
\newcommand{\omu}{{}^0\mu}              
\newcommand{\ou}{{}^0u}                 
\newcommand{\oS}{{}^0 S}                
\newcommand{\oC}{{}^0 C}                
\newcommand{\tok}{{}^0\tilde k}         
\newcommand{\tog}{{}^0\tilde g}         
\newcommand{\tg}{\tilde g}              
\newcommand{\tM}{\tilde M}              
\newcommand{\tk}{\tilde k}              
\newcommand{\tG}{\tilde G}              
\newcommand{\talpha}{\tilde \alpha}     
\newcommand{\te}{\tilde e}              
\newcommand{\ttheta}{\tilde \theta}     
\newcommand{\Ric}{\text{\rm Ric}}       
\newcommand{\SR}{{}^S R}                
\newcommand{\StR}{{}^S \tilde R}        
\newcommand{\Sg}{{}^S g}                
\newcommand{\Stg}{{}^S \tilde g}        
\newcommand{\tR}{\tilde R}              
\newcommand{\tComm}{\tilde \gamma}      
\newcommand{\tGamma}{\tilde{\Gamma}}    
\newcommand{\Comm}{\gamma}              
\newcommand{\oleq}{\preccurlyeq}        
\newcommand{\calU}{\mathcal U}
\newcommand{\calA}{\mathcal A}
\newcommand{\calF}{\mathcal F}
\newcommand{\calG}{\mathcal G}

\theoremstyle{plain}
\newtheorem{thm}{Theorem}[section]

\newtheorem{lemma}[thm]{Lemma}

\theoremstyle{remark}

\newtheorem{definition}[thm]{Definition}
\newtheorem{remark}{Remark}[section]

\numberwithin{equation}{section}        

\title{Quiescent cosmological singularities}
\author[L. Andersson]{Lars Andersson$^1$}
\address{Department of Mathematics\\
Royal Institute of Technology\\
100 44 Stock\-holm, Sweden\\
}
\email{larsa\char'100math.kth.se}

\author[A.~D. Rendall]{Alan D. Rendall$^2$}
\address{Max--Planck--Institut f\"ur Gravitationsphysik\\
Am M\"uhlenberg 1, 14424 Golm, Germany}
\email{rendall\char'100aei-potsdam.mpg.de}

\begin{document}
\date{\today \ {\em File:\jobname{.tex}}}
\begin{abstract} 
The most detailed existing proposal for the structure of spacetime
singularities originates in the work of Belinskii, Khalatnikov and
Lifshitz. We show rigorously the correctness of this proposal in
the case of analytic solutions of the Einstein equations coupled to 
a scalar field or stiff fluid. More specifically, we prove the 
existence of a family of spacetimes depending on the same number of
free functions as the general solution which have the asymptotics
suggested by the Belinskii-Khalatnikov-Lifshitz proposal near their 
singularities. In these spacetimes a neighbourhood of the singularity
can be covered by a Gaussian coordinate system in which the
singularity is simultaneous and the evolution at different spatial
points decouples. 
\end{abstract}

\maketitle
\tableofcontents

\section{Introduction}\label{sec:intro}

The singularity theorems of Penrose and Hawking are among the best known
theoretical results in general relativity. They guarantee the existence 
of spacetime singularities under rather general circumstances but say 
little about the structure of the singularities they predict. In the 
literature there are heuristic approaches to describing the structure of
singularities, notably that of Belinskii, Khalatnikov and Lifshitz (BKL),
described in \cite{lifshitz63}, \cite{bkl70}, \cite{bkl82} and their 
references. The BKL work indicates that generic singularities are 
oscillatory and therefore, in a certain sense, complicated. This complexity 
may explain why it has not been possible to determine the structure of the 
singularities by rigorous mathematical arguments.

According to the BKL analysis,
the presence of oscillatory behaviour in solutions of the Einstein equations
coupled to some matter fields is to a great extent independent of the
details of the matter content. There are, however, exceptions. It was pointed
out by Belinskii and Khalatnikov\cite{belinskii73} that a massless scalar
field can change the situation dramatically, producing singularities
without oscillations. A massless scalar field is closely related to
a stiff fluid, i.e. a perfect fluid with pressure equal to energy density,
as will be explained in more detail below. Barrow \cite{barrow78} exploited 
the singularity structure of solutions of the Einstein equations coupled to
a stiff fluid for a description of the early universe he called 
\lq quiescent cosmology\rq. We will refer to singularities where 
oscillatory behaviour is absent due to the matter content of spacetime 
as quiescent singularities.

Recently the Einstein equations coupled to a scalar field have once again 
been a source of interest, this time in the context of string cosmology.
A formal low energy limit of string theory gives rise to the Einstein
equations coupled to various matter fields. Under simplifying assumptions
the collection of matter fields can be reduced to a single scalar field,
the dilaton. The field equations are then equivalent to the standard
Einstein-scalar field equations. (Note, however, that the metric occurring
in this formulation of the equations is not the physical metric.) The
structure of the singularity in these models plays a role in the so-called 
pre-big bang scenario. For more information on these matters the reader is
referred to the work of Buonanno, Damour and Veneziano\cite{buonanno99}.

In view of the above facts, the Einstein-scalar field equations and, more 
generally, the Einstein-stiff fluid equations represent 
an opportunity to prove something about the structure of spacetime 
singularities in a context simpler than that encountered in the case
of the vacuum Einstein equations or the Einstein equations coupled to a
perfect fluid with a softer equation of state. In this paper we take
this opportunity and prove the existence of a family of solutions of the
Einstein-scalar field equations whose singularities can be described in
detail and are quiescent. These spacetimes are very general in the sense 
that no symmetry is assumed and they depend on as many free functions as the 
general solution of the Einstein-scalar field equations. They have an
initial singularity near which they can be approximated by solutions of
a simpler system of differential equations, the velocity dominated system.
Like the full system, it consists of constraints and evolution 
equations. The evolution equations contain no spatial derivatives and are
thus a system of ordinary differential equations. This is an expression
of the idea of BKL that the evolution at different spatial points decouples
near the singularity.

The structure of the paper is as follows. In the second section we recall
the Einstein-scalar field and Einstein-stiff fluid equations and define the 
corresponding velocity dominated systems. This allows the main theorems to 
be stated. They assert the existence of a unique solution of the 
Einstein-scalar field equations or Einstein-stiff fluid equations asymptotic 
to a given solution of the velocity dominated system. The proofs of the 
existence and uniqueness theorems are described in the third section. In the 
fourth section the main analytical tool used in these proofs, the theory of
Fuchsian systems, is presented. The algebraic machinery needed for the
application of the Fuchsian theory is set up in the fifth section. This 
provides the basis for the estimates of spatial curvature and other important
quantities in the section which follows. The seventh section treats relevant
aspects of the constraints. The paper concludes with a discussion of what 
can be learned from the results of the paper and what generalizations are 
desirable.

Throughout the paper the scalar field and stiff fluid cases are treated in
parallel. These are independent except in section \ref{sec:constraints}
where the propagation of constraints for the scalar field is deduced from 
the corresponding statement for the stiff fluid. Hence concentrating on
the scalar field case on a first reading would give a good idea of the main
features of the proofs.

\section{The main results}\label{sec:main}
Let $\ame_{\alpha\beta}$ be a Lorentz metric on a four-dimensional 
manifold $M$ which is diffeomorphic to $(0,T)\times S$ for a 
three-dimensional manifold $S$. Let a point of $M$ be denoted by $(t,x)$, 
where $t\in (0,T)$ and $x\in S$. It will be important in the following to 
express the geometrical quantities of interest in terms of a local frame
$\{e_a\}$ on $S$. Let $\{\theta^a\}$ denote the coframe dual to $\{e_a\}$.
Throughout the paper lower case Latin indices refer to components in this
frame, except where other conventions are introduced explicitly. Suppose 
that the metric takes the form:
\begin{equation}\label{gauss}
-dt^2+g_{ab}(t)\theta^a \otimes\theta^b
\end{equation}
where $g_{ab}(t)$ denotes the one-parameter family of Riemannian metrics 
on $S$ defined by the metrics induced on the hypersurfaces $t=$constant 
by the metric $\ame_{\alpha\beta}$. A function $t$ such that the metric
takes the form (\ref{gauss}) is called a Gaussian time coordinate.
In this case, the second fundamental form of a hypersurface 
$t=$constant is given
by $k_{ab}=-\half \d_t g_{ab}$. 
\subsection{The Einstein-matter equations}
The Einstein field equations coupled to 
matter can be written in the following equivalent $3+1$ form. The constraints 
are:
\begin{subequations}\label{constraints}
\begin{align}
R-k_{ab}k^{ab}+(\tr k)^2&=16\pi\rho\label{constrainth}             \\
\nabla^a k_{ab}-e_b(\tr k)&=8\pi j_b\label{constraintm}
\end{align}
\end{subequations}
The evolution equations are:
\begin{subequations}\label{evolution}
\begin{align}
\d_t g_{ab}&=-2k_{ab}\label{evolutiong}                        \\
\d_t k^a{}_b&=R^a{}_b+(\tr k)k^a{}_b-8\pi(S^a_{\ b}
-\half \delta^a_b\tr S)-4\pi\rho\delta^a_{\ b}\label{evolutionk}
\end{align}
\end{subequations}
Here $R$ is the scalar curvature of $g_{ab}$ and $R_{ab}$ its Ricci tensor.
The quantities $\rho$, $j_a$ and $S_{ab}$ are projections of the 
energy-momentum tensor. Their explicit forms in the cases of interest in
this paper will be given below. 

The energy-momentum tensor of a scalar field is given by
\begin{equation}\label{em}
T_{\alpha\beta}=\nabla_\alpha\phi\nabla_\beta\phi-\half (\nabla_\gamma\phi
\nabla^\gamma\phi)\ame_{\alpha\beta} 
\end{equation}
The Einstein equations can be written in the equivalent form 
$\aR_{\alpha\beta}=8\pi\nabla_\alpha\phi\nabla_\beta\phi$, where 
$\aR_{\alpha\beta}$ is the Ricci tensor of $\ame_{\alpha\beta}$.
We have $\rho=T_{00}$, $j_a=-T_{0a}$ and $S_{ab}=T_{ab}$ so that in the case
of a scalar field it follows from (\ref{em}) that:
\begin{subequations}\label{scalarem}
\begin{align}
\rho&=\half[(\d_t\phi)^2+g^{ab}e_a(\phi)e_b(\phi)]\label{scalaremrho}     \\
j_b&=-\d_t\phi e_b(\phi)\label{scalaremj}                         \\
S_{ab}&=e_a(\phi)e_b(\phi)
+\half[(\d_t\phi)^2-g^{cd}e_c(\phi)e_d(\phi)]
g_{ab}\label{scalaremS}
\end{align}
\end{subequations}
Note that it follows from the Einstein-scalar field 
equations as a consequence of
the Bianchi identity that $\phi$ satisfies the wave equation
$\ame^{\alpha\beta}\nabla_\alpha\nabla_\beta\phi=0$. This has the
$3+1$ form:
\begin{equation}\label{scalar}
-\d_t^2\phi+(\tr k)\d_t\phi+\Delta\phi=0
\end{equation}
The constraints and evolution equations are together equivalent to the
full Einstein-scalar field equations. In the following we will work with
the $3+1$ formulation of the equations rather than the four-dimensional
formulation.

A stiff fluid is a perfect fluid with pressure equal to energy density.
As we will see, it is closely related to the scalar field. The 
energy-momentum tensor of a stiff fluid is
\begin{equation}\label{stiff}
T_{\alpha\beta}=\mu (2u_\alpha u_\beta+{}^{(4)} g_{\alpha\beta})
\end{equation}
where $\mu$ is the energy density of the fluid in a comoving frame and 
$u^\alpha$ is the four-velocity. The Euler equations are obtained by
substituting this expression into the equation 
$\nabla_\alpha T^{\alpha\beta}=0$. The relation between the scalar field
and the stiff fluid is as follows. Given a solution of the Einstein
equations coupled to a scalar field, where the gradient of the scalar
field $\phi$ is everywhere timelike, define $\mu=-(1/2)\nabla_\alpha\phi
\nabla^\alpha\phi$ and $u^\alpha=\pm(-\nabla_\beta\phi
\nabla^\beta\phi)^{-1/2}\nabla^\alpha\phi$. Here the sign is chosen so
that $u^\alpha$ is future pointing, and so can be interpreted as the 
four-velocity of a fluid. Then the energy-momentum
tensor defined by (\ref{stiff}) is equal to the energy-momentum tensor
of the scalar field. Since the latter is divergence-free we see that
the fluid variables just defined together with the original metric define
a solution of the Einstein equations coupled to a stiff fluid. In the 
spacetimes of interest in the following, the gradient of $\phi$ is always 
timelike near the singularity, so that the condition
on the gradient is not a restriction in that situation. 

The matter terms needed for the Einstein equations are given in the stiff 
fluid case by
\begin{subequations}\label{stiffem}
\begin{align}
\rho&=\mu(1+2|u|^2)\label{stiffemrho}                                \\
j_b&=2\mu(1+|u|^2)^{1/2}u_b\label{stiffemj}                         \\
S_{ab}&=\mu (2u_a u_b+g_{ab})\label{stiffemS}
\end{align}
\end{subequations}
Here $|u|^2=g_{ab}u^au^b$. The Euler equations can be written in the
following 3+1 form:
\begin{subequations}\label{euler}
\begin{align}
&&\d_t\mu-2(\tr k)\mu=-2|u|^2\d_t\mu-4\mu u^a\d_t u_a-2\mu k_{ab}u^au^b
\nonumber\\
&&+2(\tr k)\mu |u|^2-2e_a(\mu) (1+|u|^2)^{1/2}u^a-\mu(1+|u|^2)^{-1/2}
u^a\nabla_a u_b u^b
\nonumber\\
&&-2\mu (1+|u|^2)^{1/2}\nabla_a u^a\label{eulermu}               \\
&&\d_t u_a+(\tr k)u_a=-\mu^{-1}[\d_t \mu-2(\tr k)\mu]u_a
+(1+|u|^2)^{-1}u^b u_a\d_t u_b\nonumber                       \\
&&-(1+|u|^2)^{-1/2}[(u_a u^b+(1/2)\delta_a^b)\mu^{-1}e_b(\mu)
+(\nabla_c u^c u_a+u^c\nabla_c u_a)]\label{euleru}
\end{align}
\end{subequations}

\subsection{The velocity dominated system}
We will prove the existence of a large class of solutions of the 
Einstein-scalar field equations and Einstein-stiff fluid equations
whose singularities we can describe in
great detail. These singularities are of the type known as velocity 
dominated. (Cf. \cite{eardley72} and \cite{isenberg90}.) This means that near
the singularity the solution can be approximated by a solution of a simpler
system, the velocity dominated system. Like the 3+1 version of the full
equations it consists of constraints and evolution equations. Solutions 
of the velocity dominated system will always be written with a left 
superscript zero and the convention is adopted that the indices of all
quantities with this superscript are moved with the velocity dominated
metric $\og_{cd}$. The velocity dominated equations will now be written out 
explicitly. The constraints are:
\begin{subequations}\label{vdconstraints}
\begin{align}
-\ok_{ab}\ok^{ab}+(\tr \ok)^2&=16\pi\orho\label{vdconstraintsh}  \\
\nabla^a (\ok_{ab})-e_b(\tr \ok)&=8\pi\oj_b\label{vdconstraintsm}
\end{align}
\end{subequations}
The evolution equations are:
\begin{subequations}\label{vdevolution}
\begin{align}
\d_t \og_{ab}&=-2\ok_{ab}\label{vdevolutiong}                   \\
\d_t \ok^a_{\ b}&=(\tr \ok)\ok^a_{\ b}-8\pi(\oS^a_{\ b}
-\half \delta^a_{\ b}\tr \oS)
-4\pi^0\rho\delta^a_{\ b}\label{vdevolutionk}
\end{align}
\end{subequations}
In these equations the matter terms are not identical to those in the full
equations but have been obtained from those by discarding certain terms. 

In the case of a scalar field the (truncated) energy-momentum tensor 
components are given by
\begin{subequations}\label{vdem}
\begin{align}
\orho&=\half(\d_t\ophi)^2\label{vdemrho}           \\
\oj_b&=-\d_t\ophi e_b(\ophi)\label{vdemj}                       \\
\oS_{ab}&=\half(\d_t\ophi)^2(\og_{ab})\label{vdemS}
\end{align}
\end{subequations}
The scalar field satisfies the equation
$$
-\d_t^2(\ophi)+(\tr \ok)\d_t\ophi=0
$$
It is important to note that the velocity dominated evolution equations 
are ordinary differential equations. However the constraints still include 
partial differential equations.

In the stiff fluid case the (truncated) energy-momentum tensor components 
are
\begin{subequations}
\begin{align}
\orho&=\omu                                                           \\
\oj_b&=2\omu \ou_b                                                   \\
\oS_{ab}&=\omu \og_{ab}
\end{align}
\end{subequations}
The velocity dominated Euler equations are
\begin{subequations}
\begin{align}
\d_t\omu-2(\tr \ok)\omu&=0                                        \\
\d_t \ou_a+(\tr \ok)\ou_a&=-(1/2)\omu^{-1}e_a(\omu)
\end{align}
\end{subequations}
In contrast to the scalar field case, this is not a system of ordinary 
differential equations. However it has a hierarchical ODE structure in
the sense that if the ODE for $\omu$ is solved and the result substituted 
into the other equations an ODE system for the $\ou_a$ results.

Substituting the expressions for the truncated energy-momentum tensor into
the velocity dominated system shows that the matter terms in the 
velocity dominated evolution equation for $\ok_{ab}$ cancel both for the
scalar field and the stiff fluid, leaving
$$\d_t \ok^a_{\ b}=(\tr \ok)\ok^a_{\ b}$$
Taking the trace of this equation gives $\d_t(\tr \ok)=(\tr \ok)^2$.
This has the general solution $\tr \ok=(C-t)^{-1}$. If we wish an initial 
singularity, as signalled by the blow-up of $\tr \ok$, to occur at $t=0$ then
$\tr \ok=-t^{-1}$. Going back to the equation for $\ok^a_{\ b}$ 
we see that $t(\ok^a_{\ b})$ is independent of time. Thus all components of the
mixed form of the second fundamental form are proportional to $t^{-1}$.

At any given spatial point we can simultaneously diagonalize $\og_{ab}$ and
$\ok_{ab}$ by a suitable choice of frame. The matrix of components of the 
metric in this frame is diagonal with the diagonal elements being 
proportional to powers of $t$. This form of the metric is that originally
used by BKL. Its disadvantage is that in general this frame cannot be 
chosen to depend smoothly on the spatial point. (There are difficulties
when there are changes in the multiplicity of the eigenvalues of 
$\ok^a_{\ b}$.) This is one reason why a different formulation is used in 
this paper. 

The velocity dominated matter equations can be solved exactly to give
\begin{equation}
\ophi(t,x)=A(x)\log t+B(x) 
\end{equation}
for given functions $A$ and $B$ on $S$ and
\begin{align}
\omu(t,x)&=A^2(x)t^{-2}        \\
\ou_a(t,x)&=t\log t (A(x))^{-1}e_a (A(x))+tB_a(x) 
\end{align}
for given quantities $A(x)$ and $B_a(x)$ on $S$.

\subsection{Statement of the main theorems}
The main theorems can now be stated.
\begin{thm}\label{maintheorem}
Let $S$ be a three-dimensional analytic manifold and let
\hfil\break
$(\og_{ab}(t),\ok_{ab}(t),\ophi(t))$ be a $C^\omega$ 
solution of the velocity dominated Einstein-scalar field equations on 
$S\times (0,\infty)$ such that $t\tr \ok=-1$ and each eigenvalue
$\lambda$ of $-t\ok^a_{\ b}$ is positive. Then there exists an open 
neighbourhood $U$ of $S\times \{0\}$ in $S\times [0,\infty)$ and a unique 
$C^\omega$ solution $(g_{ab}(t),k_{ab}(t),\phi(t))$ of the Einstein-scalar 
field equations on $U\cap (S\times (0,\infty))$ such that for each compact 
subset $K\subset S$ there are positive real numbers 
$\zeta, \beta, \alpha^a_{\ b}$, with $\zeta < \beta < \alpha^a_{\ b}$,
for which the following estimates 
hold uniformly on $K$:

\begin{enumerate}
\item \label{point:scal1st}
$\og^{ac}g_{cb}=\delta^a_{\ b}+o(t^{\alpha^a_{\ b}})$
\item
$k^a_{\ b}=\ok^a_{\ b}+o(t^{-1+\alpha^a_{\ b}})$ 
\item
$\phi=\ophi+o(t^\beta)$
\item
$\d_t\phi=\d_t\ophi+o(t^{-1+\beta})$
\item
$\og^{ac}e_f(g_{cb})=o(t^{\alpha^a_{\ b}-\zeta})$
\item \label{point:scallast}
$e_a(\phi)=e_a(\ophi)+o(t^{\beta-\zeta})$
\end{enumerate}
\end{thm}
Note that the condition on $\tr \ok$ can always be arranged by means of a
time translation and that the condition on the eigenvalues of
$\ok^a_{\ b}$ is satisfied provided it holds for a single value of $t>0$.
The positivity condition on the eigenvalues together with the velocity 
dominated Hamiltonian constraint imply that $A^2$ must be strictly
positive in the velocity dominated solution. Thus vacuum solutions are
ruled out by the hypotheses of this theorem. If an analogous analysis
were done for the Einstein equations coupled to other matter models, 
for instance a perfect fluid with equation of state $p=k\rho$, $k<1$, then  
in many cases, including that of the fluid just mentioned, the matter would 
make no contribution to the velocity 
dominated Hamiltonian constraint, so that it would not be possible to prove 
an analogous theorem. This reflects the fact that for those matter models an 
oscillatory approach to the singularity is predicted by the BKL analysis.
The scalar field is an exception, as is the stiff fluid which will be 
discussed next.

\begin{thm}\label{stiffmaintheorem}
Let $S$ be a three-dimensional analytic manifold and let 
\hfil\break
$(\og_{ab}(t),\ok_{ab}(t),\omu(t),\ou_a)$ be a $C^\omega$ 
solution of the velocity dominated Einstein-stiff fluid  equations on 
$S\times (0,\infty)$ such that $t\tr \ok=-1$ and each eigenvalue
$\lambda$ of $-t\ok^a_{\ b}$ is positive. Then there exists an open 
neighbourhood $U$ of $S\times \{0\}$ in $S\times [0,\infty)$ and and a 
unique $C^\omega$ solution $(g_{ab}(t),k_{ab}(t),\mu(t),u_a(t))$ of the 
Einstein-stiff fluid equations on $U\cap (S\times (0,\infty))$ such that 
for each compact subset $K\subset S$ there are positive real numbers 
$\zeta, \beta_1, \beta_2, \alpha^a_{\ b}$, 
with $\zeta < \beta_2 < \beta_1 < \alpha^a_{\ b}$,
for which the following estimates hold uniformly 
on $K$: 
\begin{enumerate}
\item \label{point:stiff1st}
$\og^{ac}g_{cb}=\delta^a_{\ b}+o(t^{\alpha^a_{\ b}})$
\item
$k^a_{\ b}=\ok^a_{\ b}+o(t^{-1+\alpha^a_{\ b}})$ 
\item
$\mu=\omu+o(t^{-2+\beta_1})$
\item
$u_a=\ou_a+o(t^{1+\beta_2})$
\item \label{point:stifflast}
$\og^{ac}e_f(g_{cb})=o(t^{\alpha^a_{\ b}-\zeta})$
\end{enumerate}
\end{thm}

The interest of these theorems depends very much on what information is
available on constructing solutions of the velocity dominated system.
Suppose that a solution of the velocity dominated constraints is given for
some $t=t_0>0$. The velocity dominated evolution equations constitute a system
of ordinary differential equations which can be solved with these initial
data. It follows from the remarks above that the solution
exists globally on the interval $(0,\infty)$. If we define
\begin{align}
\oC&=-\ok_{ab}\ok^{ab}+(\tr \ok)^2-16\pi \orho      \\
\oC_b&=\nabla^a (\ok_{ab})-e_b(\tr \ok)-8\pi \oj_b
\end{align}
then the velocity dominated evolution equations imply that:
\begin{align}\label{constpropab}
\d_t \oC+2t^{-1}(\oC)&=0                               \\
\d_t \oC_a+t^{-1}(\oC_a)&=\half e_a (\oC)
\end{align}
To prove this it is necessary to use the following equations for the
matter quantities, which can be derived from the velocity dominated 
matter equations in both the scalar field and stiff fluid cases.
\begin{align}
\d_t \orho&=2(\tr \ok)\orho         \\
\d_t \oj_a&=(\tr \ok)\oj_a-e_a(\orho)
\end{align}
Since $\oC$ vanishes at $t=t_0$ the evolution equation for $\oC$ implies 
that it vanishes everywhere. Then the evolution equation for $\oC_a$, 
together with the fact that it vanishes for $t=t_0$ implies that $\oC_a$ 
vanishes everywhere.
To sum up, if the velocity dominated constraints are satisfied at some 
time and the velocity dominated evolution equations are satisfied everywhere
then the velocity dominated constraints are satisfied everywhere. Thus in 
order to have a parametrization of the general solution of the velocity 
dominated system, it is enough to obtain a parametrization of solutions 
of the velocity dominated constraints. The latter question will be treated 
in section \ref{sec:constraints}.

\section{Framework of the proofs}\label{sec:framework}

In this section the proofs of Theorems \ref{maintheorem} and 
\ref{stiffmaintheorem} are outlined. Only the general logical
stucture of the proof is explained here and the hard technical parts
of the argument are left to later sections. These results will be referred
to as required in this section.

The first step is to make a suitable ansatz for the desired solution. This
essentially means giving names to the remainder terms occurring in the
statements of the main theorems. Assume that a velocity dominated solution
is given as in those statements. Then a solution is sought in the form:

\begin{subequations}\label{ansatz-intro}
\begin{align}
g_{ab}&=\og_{ab}+\og_{ac}t^{\alpha^c_{\ b}}\gamma^c_{\ b} 
\label{ansatzg}     \\
k_{ab}&=g_{ac}(\ok^c_{\ b}+t^{-1+\alpha^c_{\ b}}\kappa^c_{\ b})
\label{ansatzk}   
\end{align}
\end{subequations}
In the following the summation convention applies only to repeated tensor
indices and not to non-tensorial quantities like $\alpha^a_{\ b}$. Thus in the
above equations there is a summation on the index $c$ but none on the index 
$b$. Matter fields are sought in the form:
\begin{equation}\label{ansatzp-intro}
\phi=\ophi+t^{\beta}\psi
\end{equation}
and
\begin{subequations}\label{stiffansatz-intro}
\begin{align}
\mu&=\omu+t^{-2+\beta_1}\nu\label{ansatzmu-intro}                 \\
u_a&=\ou_a+t^{1+\beta_2}v_a
\end{align}
\end{subequations}
respectively. The Einstein-scalar field equations (\ref{constraints}), 
(\ref{evolution}), (\ref{scalarem}) and (\ref{scalar}) can be rewritten as 
equations for $\gamma^a_{\ b}$, $\kappa^a_{\ b}$ and $\psi$. Similarly the 
Einstein-stiff fluid equations can be written as equations for
$\gamma^a_{\ b}$, $\kappa^a_{\ b}$, $\nu$ and $v_a$. This system of equations
(for either choice of matter model) will be called the 
{\bf first reduced system}.
Since $\gamma^a_{\ b}$ and $\kappa^a_{\ b}$ 
are mixed tensors, there is no direct 
way to express the fact that they originated from symmetric tensors. Instead 
it must be shown that when the first reduced system is solved with 
suitable asymptotic conditions as $t\to 0$ then the quantities $g_{ab}$ and 
$k_{ab}$ defined by the above equations are in fact symmetric as a 
consequence of the differential equations and the initial conditions.
When allowing non-symmetric tensors $g_{ab}$ and $k_{ab}$, we need to
establish some conventions in order to make the definition of the
first reduced system unambiguous. Firstly, define $g^{bc}$ as the
unique tensor which satifies $g_{ab}g^{bc}=\delta^{\ c}_a$. Next, use the
convention that indices on tensors are lowered by contraction with the
second index of $g_{ab}$ and raised with the first index of $g^{bc}$.
This maintains the usual properties of index manipulations in the case
of symmetric $g_{ab}$ as far as possible. The covariant derivatives in
the equations are expressed in terms of the connection coefficients in
the frame $\{e_a\}$ in an unambiguous way. The definition of the 
connection coefficients is extended to the case of a non-symmetric
tensor $g_{ab}$ by fixing the order of indices according to
\begin{equation}\label{eq:conncoeff}
g_{cd} \Gamma_{ab}^d = \half ( 
e_a(g_{bc}) + e_b(g_{ac})
- e_c(g_{ab}) + \Comm_{ab}^d g_{cd} - \Comm_{ac}^d g_{bd} - \Comm_{bc}^d g_{ad}
)
\end{equation}
where $\Comm_{ab}^c = \theta^c([e_a, e_b])$ are the structure functions 
of the frame. 

Finally, in order to define the Ricci tensor in the
evolution equation for $k_{ab}$ we define $R_{ab}$ to be the Ricci tensor
of the symmetric part $\Sg_{ab}=(1/2)(g_{ab}+g_{ba})$ of $g_{ab}$. 
In Lemma \ref{symmetry} it is shown that given a solution of the velocity 
dominated system as in the statement of one of the main theorems
\ref{maintheorem}--\ref{stiffmaintheorem}, any
solution of the first reduced system which satisfies points
\ref{point:scal1st}--\ref{point:scallast} of Theorem \ref{maintheorem}, in
the scalar field case, and
points \ref{point:stiff1st}--\ref{point:stifflast} of Theorem
\ref{stiffmaintheorem}, in the stiff fluid case, 
gives rise to symmetric tensors $g_{ab}$ and $k_{ab}$, and thus 
to a solution of the Einstein-matter evolution equations. It then follows from
Lemma \ref{smallconstraints} and the remarks following it that the 
Einstein-matter constraints are also satisfied. It follows that to prove the 
main theorems it is enough to prove the existence and uniqueness of solutions 
of the first reduced system of the form given in the main theorems.

The existence theorem for the first reduced system will be proved using the
theory of Fuchsian systems. Since the form of this theory we use in the
following concerns a system of first order equations it is not immediately
applicable, due to the occurrence of second order derivatives of the 
metric and scalar field. It is necessary to introduce some suitable new 
variables representing spatial derivatives of the basic variables.
Define $\lambda^a_{bc}=t^\zeta e_c(\gamma^a_{\ b})$ and, in the scalar field 
case, $\omega_a=t^\zeta e_a(\psi)$ and $\chi=t\d_t\psi+\beta\psi$, where 
$\zeta$ and $\beta$ are positive constants. The evolution equations 
satisfied by these quantities are given explicitly in (\ref{einsteinnslam})
and (\ref{eq:wave-first}). With the help of the new variables these equations 
together with the first reduced system can be written as a first order 
system. Call the result the {\bf second reduced system}. 
It is easy to show, using the evolution
equations for differences like $\gamma^a_{bc}-t^\zeta e_c (\gamma^a_{\ b})$
which follow from the second reduced system, that the first and second reduced 
systems give rise to the same sets of solutions under the assumptions of the 
main theorems, together with corresponding assumptions on the new variables.
Thus it suffices to solve the second reduced system, which is of first order. 

If it can be shown that the second reduced system is Fuchsian, then the
main theorems follow from Theorem \ref{fuchstheorem}. In fact it is enough 
to show that the restriction of the system to a neighbourhood of an 
arbitrary point of $S$
is Fuchsian. For the local solutions thus obtained can be pieced together
to get a global solution. Moreover asymptotic estimates as in the
statements of the main theorems follow from corresponding local statements,
since a compact subset of $S$ can be covered by finitely many of the local
neighbourhoods.

In sections \ref{sec:setup} and \ref{sec:curvest} it is shown that the second 
reduced system is Fuchsian on some neighbourhood of each point of $S$ for
a suitable choice of the constants $\alpha^a_{\ b}$, $\beta$, $\beta_1$,
$\beta_2$ and $\zeta$ depending on 
the given velocity dominated solution. This requires a detailed 
analysis of the degree of singularity of all terms in the second reduced
system, in particular that of the Ricci tensor. The result of the latter is
Lemma \ref{lem:curvest}. This is the hardest part
of the proof. Due the above considerations, it is clear that the main
theorems follow directly from Theorem \ref{fuchstheorem}.

\section{Fuchsian systems}\label{sec:fuchs}
The proofs of Theorems \ref{maintheorem} and \ref{stiffmaintheorem} rely on a 
result of Kichenassamy 
and Rendall\cite{kichenassamy98} on Fuchsian systems which uses a
method going back to Baouendi and Goulaouic\cite{baouendi77}. The result of 
\cite{kichenassamy98} will now be recalled. It concerns a system of the form:
\begin{equation}\label{fuchs}
t\frac{\d u}{\d t}+A(x)u=f(t,x,u,u_x)
\end{equation}
Here $u(t,x)$ is a function on an open subset of $\R\times\R^n$ with values 
in $\R^k$ and $A(x)$ is a $C^\omega$ matrix-valued function. The derivatives 
of $u$ with respect to the $x$ variables are denoted by $u_x$. The function 
$f$ is defined on $(0,T_0]\times U_1\times U_2$, where $U_1$ is an open subset 
of $\R^n$ and $U_2$ is an open subset of $\R^{k+nk}$, and takes values in 
$\R^k$. We assume that $A(x)$ is defined on $U_1$. 

In this and later sections it will be useful to have some terminology for
comparing the sizes of certain expressions.
\begin{definition}\label{def:oleqdef}
Let $F(t,x,p)$, $G(t,x,p)$ be functions on $(0,T_0] \times U_1 \times U_2$, 
where $U_1, U_2$ 
are open subsets of $\R^n$ and $\R^N$ respectively. Then we will say that 
$$
F \oleq G 
$$
if for every compact $K \subset U_1 \times U_2$, there is a constant $C$ such
that 
$$
|F(t,x,p)| \leq C|G(t,x,p)| \quad \text{for $t \in (0,t_0]$, $(x,p)\in K$}.
$$
\end{definition}
In the particular case that $G$ is just a function of $t$ we will often use 
the familiar notation $F=O(G(t))$ to replace $F\oleq G$. The notation
$F=o(G(t))$ will also be used to indicate that $F/G$ tends to zero uniformly
on compact subsets of $U_1\times U_2$ as $t\to 0$.   

In the theorem on Fuchsian systems the function $f$ is supposed to be 
regular in a sense which will now be explained. To do this we need the notion
of a function which is continuous in $t$ and analytic in other (complex) 
variables. This means by definition that it should be a continuous function 
of all variables, that the first order partial derivatives with respect to 
all variables other than $t$ should exist and be continuous, and that the 
Cauchy-Riemann equations should be satisfied in these variables. For further 
remarks on this concept see \cite{rendall90}. Assume that there is an open 
subset $\tilde U$ of $\C^{n+k+nk}$ whose intersection with the real section 
is equal to $U_1\times U_2$ and a function $\tilde f$ on
$(0,T_0]\times\tilde U$ continuous in $t$ and analytic in the remaining 
arguments whose restriction to $(0,T_0]\times\R^{n+k+nk}$ is equal to $f$. 
The function $f$ is called {\bf regular} if it has an analytic continuation 
$\tilde f$ of the kind just described, and if there is some $\theta>0$ such 
that $\tilde f$ and its first derivatives with respect to the arguments $u$ 
and $u_x$ are $O(t^\theta)$ as $t\to 0$, in the sense introduced above.

For a matrix $A$ with entries $A^a_{\ b}$ let $\|A\|=\sup \{\|Ax\|:\|x\|=1\}$
(operator norm) and $\|A\|_\infty = \max_{a,b}|A^a_{\ b}|$ (maximum norm).
Since these two norms are equivalent the operator norm could be replaced by
the maximum norm in the statement of the theorem which follows and in Lemma
\ref{matrix} below. However in the proof of that lemma  below the use of the 
operator norm is important.
\begin{thm}\label{fuchstheorem}
Suppose that the function $f$ is regular, $A(x)$ has an analytic continuation 
to an open set $\tilde U_1$ whose intersection with the real section is $U_1$, 
and there is a constant $C$ such that $\|\sigma^{A(x)}\|\le C$ for 
$x\in\tilde U_1$ and $0<\sigma<1$. Then the equation (\ref{fuchs}) has a 
unique solution $u$ defined near $t=0$ which is continuous in $t$ and 
analytic in $x$ and tends to zero as $t\to 0$. If $\tilde f$ is analytic for 
$t>0$ then this solution is also analytic in $t$ for $t>0$.
\end{thm}

\begin{remark}
Under the hypotheses of the theorem spatial derivatives of any order of $u$
are also $o(1)$ as $t\to 0$. This follows directly from the proof of the
theorem in \cite{kichenassamy98}.
\end{remark} 

\begin{remark} 
If the coefficients of the equation depend analytically on a parameter and
are suitably regular, then the solution depends analytically on the 
parameter. It suffices to treat the parameter as an additional spatial
variable.
\end{remark}

The statement of the theorem is not identical to that given in
\cite{kichenassamy98} but the proof is just the same. We can write
$f(t,x,u,u_x)=t^\theta g(t,x,u,u_x)$ for some bounded function $g$.
By replacing $t$ as time variable by $t^\theta$, it can be assumed
without loss of generality that $\theta=1$. Then the iteration used
in the proof given in \cite{kichenassamy98} converges to the desired
solution under the regularity hypothesis we have made on $f$. 

For the applications in this paper an extension of this result which 
applies to equations slightly more general than (\ref{fuchs}) will be 
required. These have the form
\begin{equation}\label{fuchst}
t\frac{\d u}{\d t}+A(x)u=f(t,x,u,u_x)+g(t,x,u)t\frac{\d u}{\d t}
\end{equation} 
If we have an equation of this form where $f$ and $g$ are regular
then an analogous existence and uniqueness result holds. For we can 
rewrite (\ref{fuchst}) in the form
\begin{equation}
t\frac{\d u}{\d t}+A(x)u=[I-(I-g(t,x,u))^{-1}]A(x)+(I-g(t,x,u))^{-1}
f(t,x,u,u_x)
\end{equation}
If we call the right hand side of this equation $h(t,x,u,u_x)$ then it
satisfies the conditions required of $f(t,x,u,u_x)$ in Theorem
\ref{fuchstheorem}. In other words $h$ is regular. For if $g(t,x,u)$ is 
$O(t^\theta)$ then $(I-g(t,x,u))^{-1}$ is $O(1)$ and 
$[I-(I-g(t,x,u))^{-1}]$ is $O(t^\theta)$.

In the following it will be necessary to verify the regularity hypothesis for
some particular systems, and some general remarks which can be used to 
simplify this task will now be made. In these systems we always have
\begin{equation}
f(t,x,u,u_x)=\sum_{i=1}^m t_i F_i(x,v(t,x),u,u_x)
\end{equation}
where the $F_i$ are analytic functions on an open subset $V$ of 
$\R^{n+l+k+nk}$ which includes $U_1\times\{0\}\times U_2$,
$t_1,\ldots,t_m$ are some functions
of $t$ and $v(t,x)$ is a given function with values in $\R^l$. The 
$t_i$ are continuous functions on $(0,T_0]$ which tend to zero as $t\to 0$
as least as fast as some positive power of $t$ and are analytic for $t>0$. 
The function $v(t,x)$ is the restriction of an analytic function on 
$(0,T_0]\times\tilde U_1$ to real values of its arguments. Each component 
of $v$ tends to zero as $t\to 0$, uniformly on $\tilde U_1$. These properties
ensure that $f$ is regular. For the functions $F_i$ have analytic 
continuations to some open neighbourhood $\tilde V$ of $V$ in 
$\C^{n+l+k+nk}$.

In the examples we will meet the functions $t_i$ are either positive 
powers of $t$ or positive powers of $t$ times positive powers of $\log t$.
The role of $v(t,x)$ is played by the functions $t_j$, the components of
the velocity dominated metric $\og_{ab}$, their spatial derivatives of 
first and second order, and the components of the inverse metric
multiplied by suitable powers of $t$ so that the product vanishes in
the limit $t\to 0$. We saw in the last section that 
$\og_{ab}$ can be written in the form $t^{2K}$ which, for each fixed $t$, is 
an entire function of $K$. Thus $t^{2K(x)}$ is analytic on any region where 
$K(x)$ is analytic. The velocity dominated metric and its derivatives all 
tend to zero uniformly on compact sets as $t\to 0$ while the same is true 
of the inverse metric multiplied by a suitable power of $t$.

Next a criterion will be given which allows the hypothesis on the matrix 
$A$ in Theorem \ref{fuchstheorem} to be checked in many cases.
\begin{lemma}\label{matrix}
Let $A(x)$ be a $k\times k$ matrix-valued continuous function defined on 
a compact subset of $\R^n$. If there is a constant $\alpha$ such that, for 
each eigenvalue $\lambda$ of $A(x)$ at any point of the given compact set,
${\rm Re}\lambda>\alpha$ then there is a constant $C$ such that the
estimate $\|t^{A(x)}\|\le Ct^\alpha$ holds for $t$ small and
positive and $x$ in the compact set.
\end{lemma}

\noindent
{\bf Proof} The general case can be reduced to the case $\alpha=0$ by
the following computation:
\begin{equation}
\|t^{A-\alpha I}\|\le\|t^A\|t^{-\alpha}
\end{equation}
In the rest of the proof only the case $\alpha=0$ will be considered.
Without the parameter dependence the result could easily be proved by 
reducing the matrix to Jordan canonical form. The difficulty with a 
parameter is that the reduction to canonical form is in general not a 
continuous process. At least it can be concluded from the continuity 
properties of the eigenvalues that there is a $\beta>0$ such that
all eigenvalues satisfy ${\rm Re}\lambda>\beta$. Let $s=\log t$.
Then the problem is to show that for a fixed $s_0$ there is a constant 
$C>0$ such that $\|e^{-sA}\|\le C$ for all $s>s_0$. By scaling $t$ we may 
suppose without loss of generality that $s_0=0$. For each $x$ we
can conclude by reduction to canonical form that exists a value
$s_x$ of $s$ such that the inequality 
$\|e^{-s_xA(x)}\|<e^{-\beta s_x/2}$ holds. By continuity of the exponential 
function there is an open neighbourhood $U_x$ of $x$ where this continues 
to hold for the given value of $s_x$. Let 
$C_x=\sup\{\|e^{-sA(y)}\|:s\in [0,s_x],y\in U_x\}$.
It follows that for any $s\in [0,\infty)$ and any $y\in U_x$ we have
\begin{equation}
\|e^{-sA(y)}\|\le C_x\|e^{-s_xA(y)}\|^{[s/s_x]}\le C_x
e^{-\beta s_x[s/s_x]/2}
\le C_x 
\end{equation}
By compactness, it is possible to pass to a subcover consisting of a
finite number of the sets $U_x$ and letting $C$ be the maximum of the
corresponding $C_x$ we obtain the required estimate.

\begin{remark} \label{lem:matrix-remark}
More generally, an analogous estimate is obtained if $A(x)$ is the direct 
sum of a matrix $B(x)$ whose eigenvalues have positive real parts and the
zero matrix. This is obvious since in that case $t^{A(x)}$ is the direct sum
of $t^{B(x)}$ and the identity.
\end{remark}

\section{Setting up the reduced equations}\label{sec:setup}
In this section we introduce the adapted frame $\{e_a\}$ and the auxiliary
exponents $\{q_a\}$ which will be used in the curvature estimate. 

Let $x_0$ be given. Let $\og_{ab}$, $\ok^a_{\ b}$,  be solutions of the
velocity dominated evolution and constraint equations (\ref{vdevolution})
and (\ref{vdconstraints}). Let $p_a$ be the eigenvalues of 
$K^a_{\ b} = -t \ok^a_{\ b}$. 
Assume that $\{p_a\}$ are such that 
$p_a(x_0) > 0$, $a=1,2,3$, $\sum_a p_a = 1$ (Kasner condition) and $p_a$ are
ordered so that $p_a \leq p_b$, for  $a \leq b$.
Fix an initial 
time $t_0 \in (0,1)$. We will in the following restrict our considerations
to $t \in (0,1)$. 
 
If $K$ has a double eigenvalue at $x_0$, then in general the eigenvalues and
eigenvectors of $K$ are not analytic in a neighbourhood of $x_0$, 
and therefore in
general it is not possible to introduce an analytic frame diagonalizing $K$
in a given neighbourhood. We will avoid this problem by using the well known
fact that if $p_{a'}$ is a double eigenvalue of $K(x_0)$, the eigenspace of
the pair of eigenvalues corresponding to $p_{a'}$ is analytic in a
neighbourhood of $x_0$. This means in particular that it is possible to choose
an analytic frame which is adapted to the eigenspace of the pair of
eigenvalues. This will play a central role in what follows. 

Choose numbers $\alpha_0, \eps > 0$ so that
$\eps = \alpha_0/4  < \min\{p_a(x_0)\}/ 40$.
\begin{enumerate}
\item \label{point:cases} 
{\bf Cases I, II, III:}
We will distinguish between the following cases:
\begin{enumerate}\renewcommand{\theenumii}{{\Roman{enumii}}}
\item (near Friedmann) $\max_{a,b}|p_a - p_b| < \eps/2
, a=1,2,3$. 
\item (near double eigenvalue) $\max_{a,b}|p_a - p_b| >
\eps/2$, and $|p_{'a} - p_{b'}| < \eps/2$ for some pair $a',b'$, $a' \ne b'$. 
Denote by
$p_{\perp}$ the distinguished exponent not equal to $p_{a'}, p_{b'}$.
\item (diagonalizable) $\min_{\substack{a,b\\a \ne b}}|p_a - p_b| > \eps/2$
\end{enumerate}
By reducing $\eps$ if necessary we can make sure that condition I,
II or III holds at $x_0$ if the maximum multiplicity of an eigenvalue at
$x_0$ is three, two or one respectively.
The conditions I, II, III are open, and hence there is an open
neighbourhood $U_0 \ni x_0$ such that, if condition I, II or III holds at
$x_0$, the condition holds in $U_0$ and further, for $x\in U_0$, 
$\min_a \{p_a(x)\} > 20\eps$. 
\item \label{point:exponents}
{\bf Auxiliary exponents $\{q_a\}$:}
Let $U_0 \ni x_0$ be as in point \ref{point:cases}. 
We will define analytic functions 
$q_a, a=1,2,3$, called auxiliary exponents, in 
$U_0$, with the properties
\begin{enumerate}
\item $q_a > 0$ (positivity)
\item $q_a \leq q_b$ if $a \leq b$ (ordering) 
\item $\sum_a q_a =1  $ (Kasner)
\end{enumerate}
The auxiliary exponents $q_a$ will be defined in terms of the
eigenvalues $p_a$ of $K$ depending on whether at $x_0$ we are in
case I, II, or III.

Let $q_a(x_0) = p_a(x_0)$, $a = 1,2,3$, and choose $q_a$ on 
$U_0$ satisfying the positivity, ordering and
Kasner conditions such that in cases I, II, III, the following holds. 
\begin{enumerate}\renewcommand{\theenumii}{{\Roman{enumii}}}
\item $q_a = 1/3$, $a=1,2,3$
\item $q_{\perp} = p_{\perp}$, $q_{a'} = q_{b'} =
\half(1-q_{\perp})$
\item $q_a = p_a$
\end{enumerate}
Then $q_a$ are analytic on $U_0$. Note that it follows from the definition
of $q_a$ that $q_1 \geq \min_i\{p_i\}$ and that 
$\max_a |q_a - p_a| < \eps/2$. 
\item \label{point:frame}
{\bf The frame $\{e_a\}$:}
In each case I, II, III
define an analytic frame 
$\{e_a\}$ with dual frame $\{\theta^a\}$, by the following prescription:
$\{e_a\}$ is an ON frame w.r.t. $\og(t_0)$,
and in case  II, III the
following additional conditions hold.
\begin{enumerate}\renewcommand{\theenumii}{{\Roman{enumii}}} \setcounter{enumii}{1}
\item $e_{\perp}$ is the eigenvector of $K$ corresponding to 
$q_{\perp}$ and 
$e_{a'}, e_{b'}$ span the eigenspace of $K$ corresponding to
the eigenvalues $p_{a'}, p_{b'}$. 
\item $e_a$ are 
eigenvectors of $K$ corresponding to the eigenvalues $q_a$. 
\end{enumerate}
\end{enumerate} 
We will call $\{q_a\}$, $\{e_a\}$, $\{\theta^a\}$, 
satisfying the above conditions {\bf
adapted}. 
In the following we will work in a neighbourhood $U_0$ defined as
above and assume 
that we are given adapted $\{q_a\}$, $\{e_a\}$, $\{\theta^a\}$, on $U_0$.

The role of $\alpha^a_{\ b}$ will be to shift the spectrum of the system
matrix to be positive. We need to choose $\alpha_0$ larger than $\eps$ to
compensate for the fact that the $q_a$ are not the exact eigenvalues of
$K$. 

In the following $T_{ab}$ will denote frame components
of the tensor $T$ w.r.t. the frame $\{e_a\}$.
Define the rescaled frame $\te_a = t^{-q_a} e_a$ with dual frame $\ttheta^a =
t^{q_a} \theta^a$ and denote by $\tilde T_{ab}$ the $\te_a$ frame components of
the tensor $T$. Then we have 
\begin{equation}\label{eq:resc}
\tilde T_{ab} = t^{-q_a - q_b} T_{ab} , \qquad 
\tilde T^a_{\ b} = t^{q_a - q_b} T^a_{\ b} . 
\end{equation}

It follows from the definitions, that in case III, 
\begin{equation}\label{eq:IIIdiag}
K^a_{\ b} = \delta^a_{\ b} q_b , \qquad 
\ok^a_{\ b} = - t^{-1} \delta^a_{\ b} q_b
\end{equation}
while in case II, the tensors 
$K$, $\ok$ and $\og$ are block diagonal in the frame
$\{e_a\}$, 
\begin{equation}\label{eq:IIdiag}
K^{a'}_{\ \perp} = 0, \quad \ok^{a'}_{\ \perp} = 0 , \quad \og_{a' \perp} = 0
\end{equation}

For $s \in \R$, let $(s)_+ = \max(s,0)$, for $a,b \in \{1,2,3\}$, let 
\begin{equation}\label{eq:alphadef}
\alpha^a_{\ b} = 2(q_b - q_a)_+ + \alpha_0 ,
\end{equation}
and 
$\talpha^a_{\ b} = |q_b - q_a| + \alpha_0$. In view of the relation 
$
2(s)_+ -s = |s|
$
we have 
\begin{equation}\label{eq:talpha}
\alpha^a_{\ b} + q_a - q_b = \talpha^a_{\ b} .
\end{equation}
 
The following identities are an immediate consequence of equations
(\ref{eq:IIIdiag}) and (\ref{eq:IIdiag}), together with the fact that in
case I, $q_a = 1/3$, $a=1,2,3$.
\begin{subequations}\label{eq:VDrel}
\begin{align}
\og_{ab} t^{\alpha^b_{\ c}}  &= \og_{ab} t^{\alpha^a_{\ c}}
 \label{eq:galpha}
\\
\og_{ab} t^{q_b} &= \og_{ab} t^{q_a} \label{eq:gq}
\end{align}
\end{subequations}
Note that in (\ref{eq:VDrel}) no summation over indices is implied. 

Lemma \ref{matrix} implies the following estimate for the rescaled frame
components of $\tog$. 
\begin{lemma}\label{lem:gest}
\begin{equation}\label{eq:gest}
||\tog_{ab}||_\infty\leq Ct^{-\eps}, \qquad ||\tog^{ab}||_\infty 
\leq C t^{-\eps}
\end{equation}
\end{lemma}
\begin{proof}
Let $Q^a_b = \delta^a_{\ b} q_b$. A direct computation starting from the
matrix form of the velocity dominated evolution equation gives 
$$
t\d_t \tG_{ab} = 2 \tG_{ac}(K^c_{\ b} - Q^c_{\ b})
$$
which has the solution 
$$
\tG_{ab}(t) = \tG_{ac}(t_0) \left ( \frac{t}{t_0} \right )^{2(K^c_{\ b} -
Q^c_{\ b})}
$$

From the definition of $q_a$ and
the frame $e_a$ the spectrum of $K-Q$ is of the form $p_a - q_a$. Therefore
since $|p_a - q_a|< \eps/2$, we find that the spectrum of $2(K-Q)$ is
contained in the interval $(-\eps, \eps)$.
Therefore it follows from Lemma \ref{matrix} that 
$||\tG||_\infty\leq Ct^{-\eps}$. 

Similarly, using the fact that 
$\tG^{-1}$ satisfies the equation 
$$
t\d_t \tG^{-1}_{ab}  = 2 (Q_a^{\ c} -K_a^{\ c})\tG^{-1}_{cb}
$$
an application of Lemma \ref{matrix} yields the estimate 
$||\tG^{-1}||_\infty \leq C t^{-\eps}$.
\end{proof}

We are now ready to describe the ansatz which will be used to write the
Einstein--scalar field and Einstein--stiff fluid systems in Fuchsian
form. Assume that a 
solution of the 
velocity dominated constraint and evolution equations, $\og_{ab}$,
$\ok_{ab}$,
with adapted frame,
coframe, and auxiliary exponents $\{e_a\}$, $\{\theta^a\}$, $\{q_a\}$, 
is given. Let $\alpha^a_{\ b}$ be defined by (\ref{eq:alphadef}). 

Let $\zeta = \eps/200$. 
We will consider metrics and second fundamental forms 
$g,k$ of the form
\begin{subequations}\label{eq:ansatz}
\begin{align}
g_{ab} &= \og_{ab} + \og_{ac} t^{\alpha^c_{\ b}} \gamma^c_{\ b} &&
\gamma^c_{\ b} = \ordo(1) \label{eq:gdef} \\
g^{ab}&= \og^{ab} + t^{\alpha^a_{\ b}} \bar \gamma^a_{\ c} \og^{cb} &&
\bar \gamma^a_{\ c} = \ordo(1) \label{eq:ginvdef} \\ 
e_c(\gamma^a_{\ b}) &= t^{-\zeta} \lambda^a_{bc} && \lambda^a_{bc} =
\ordo(1) \label{eq:lambda-ansatz} \\
k_{ab} &= g_{ac} ( \ok^c_{\ b} + t^{-1+\alpha^c_{\ b}} \kappa^c_{\ b} )
&& \kappa^c_{\ b} = \ordo(1) 
\end{align}
\end{subequations}
The form (\ref{eq:ginvdef}) is a consequence of (\ref{eq:gdef}). To see
this, note that 
$
\og^{ac} g_{cb} = \delta^a_{\ b} + t^{\alpha^a_{\ b}} \gamma^a_{\ b}
$
and that 
$
g^{ac} \og_{cb} = (\og^{ac} g_{cb} )^{-1}. 
$
Thus the desired result follows from the matrix identity 
$
(I+A)^{-1}=I-A+(I+A)^{-1}A^2
$
and the fact that, using $2(x)_+ - x = |x|$, it can be concluded that 
\begin{equation}\label{alphacombo}
\alpha^a_{\ e} + \alpha^e_{\ b} - \alpha^a_{\ b} = 
|q_e - q_a| + |q_b - q_e| - |q_b - q_a| + \alpha_0 \geq \alpha_0 .
\end{equation}
The latter relation shows that each component of the square of $\gamma^a{}_b$ 
vanishes faster than the corresponding component of $\gamma^a{}_b$ itself.

Let $\beta = \eps/100$.
In addition to (\ref{eq:ansatz}) we will use the following ansatz for the 
scalar field
\begin{subequations}\label{eq:ansatzp}
\begin{align}
\phi &= \ophi+t^{\beta}\psi && \psi = \ordo(1) \\
e_a(\psi) &= t^{-\zeta} \omega_a && \omega_a = \ordo(1)
\label{eq:omega-ansatz}
\\ 
t\d_t\psi+\beta\psi &= \chi && \chi = \ordo(1) 
\end{align}
\end{subequations}
and for the stiff fluid case, 
\begin{subequations}\label{eq:stiffansatz}
\begin{align}
\mu&=\omu+t^{-2+\beta_1}\nu && \nu = \ordo(1) \label{ansatzmu}                 \\
u_a&=\ou_a+t^{1+\beta_2}v_a && v_a = \ordo(1)
\end{align}
\end{subequations}

Equations (\ref{eq:ansatz}), (\ref{eq:ansatzp}) and (\ref{eq:stiffansatz}) 
with the exception of (\ref{eq:lambda-ansatz}) and (\ref{eq:omega-ansatz}) 
will be used to derive the first reduced form of the field equations, and 
equations (\ref{eq:lambda-ansatz}) and (\ref{eq:omega-ansatz}) for the 
spatial derivatives of $\gamma^a_{\ b}$ and $\psi$, will be used to
derive the second reduced system.

Note that in view of (\ref{eq:resc}), we have
\begin{subequations}\label{eq:tansatz}
\begin{align}
\tg_{ab} &= \tog_{ab} + \tog_{ac} t^{\talpha^c_{\ b}} \gamma^c_{\ b} \\
\tg^{ab} &= \tog^{ab} + t^{\talpha^a_{\ b}} \bar \gamma^a_{\ c} \tog^{cb} 
\label{eq:tgup} \\
\tk_{ab} &= \tg_{ac} ( \tok^c_{\ b} + t^{-1+\talpha^c_{\ b}} \kappa^c_{\ b})
\end{align}
\end{subequations}

We use the following {\bf conventions} throughout:
\begin{itemize}
\item 
indices on velocity dominated
fields $\og_{ab}, \ok_{ab}, \ou_a$ are raised and 
lowered with $\og_{ab}$, while indices on other
tensors are raised and lowered with $g_{ab}$. 
\item the dynamic tensor fields $\gamma^a_{\ b}, \kappa^a_{\ b}$ in 
$g_{ab}, k_{ab}$ are always
used in mixed form and only in $\{e_a\}$ frame components. 
\item the dynamic 1--form $v_a$ in the velocity field $u_a$ is always used
with lower index.
\end{itemize}

\subsection{The reduced Einstein--matter system}
In this section, we describe the first reduced system for the
Einstein--scalar field evolution equations, derived from (\ref{evolution})
using the ansatz given by equations (\ref{eq:ansatz}) and (\ref{eq:ansatzp}) 
for $g_{ab}, k_{ab}, \phi$ in terms of the velocity dominated solution 
$\og_{ab}, \ok_{ab}, \ophi$, the auxiliary exponents $\{q_a\}$ and the 
dynamical fields $\gamma^a_{\ b}, \kappa^a_{\ b}, \psi$. Similarly, we
describe the first reduced system for the Einstein--stiff fluid evolution
equations obtained using the equations (\ref{eq:stiffansatz}). For 
convenience we use the term \lq Einstein--matter system\rq\ to describe the 
Einstein--scalar field and Einstein--stiff fluid system collectively.

The tensor $g_{ab}$ of the form (\ref{eq:gdef}) is not a priori symmetric, 
but it will follow from Lemma \ref{symmetry} that the solution to
the Fuchsian form of the 
Einstein--matter evolution equations will be symmetric. It is
convenient to introduce the symmetrized tensor 
$$
\Sg_{ab} = \half ( g_{ab} + g_{ba} ) .
$$
Let $\SR_{ab}$ be the Ricci tensor computed w.r.t. the symmetrized metric
${}^S g_{ab}$, see section \ref{sec:curvest} for details.

By substituting into the evolution equations (\ref{evolution}) with $R_{ab}$
replaced by $\SR_{ab}$, defined in terms of $\gamma^a_{\ b}$ and 
$\lambda^a_{\ bc}$,
we get the following system for 
$\gamma^a_{\ b}, \kappa^a_{\ b},\lambda^a_{\ bc}$.
\begin{subequations}\label{eq:ein-first-red}
\begin{align}
t\d_t \gamma^a{}_b&+\alpha^a_{\ b} \gamma^a_{\ b}+2\kappa^a_{\ b}
+2\gamma^a_{\ e}(t\ok^e_{\ b})-2(t\ok^a_{\ e})\gamma^e_{\ b}=
-2t^{\alpha^a_{\ e} + \alpha^e_{\ b} 
- \alpha^a_{\ b}} \gamma^a_{\ e}\kappa^e_{\ b}
\label{einsteinnsg} \\ 
t\d_t\lambda^a_{bc}&=t^\zeta e_c (t\d_t\gamma^a_{\ b})+\zeta t^{\zeta}
e_c(\gamma^a_{\ b}) \label{einsteinnslam}\\
t\d_t\kappa^a_{\ b}&+\alpha^a_{\ b}\kappa^a_{\ b}
-(t\ok^a_{\ b})(\tr\kappa)
=t^{\alpha_0}(\tr\kappa)\kappa^a_{\ b} \\
&\quad
+t^{2-\alpha^a_{\ b}} ( \SR^a_{\ b}- M^a_{\ b} )
\label{einsteinnsk}
\end{align}
\end{subequations}
where $M_{ab}$ is is given by  
\begin{subequations}\label{eq:Mform}
\begin{align}
M_{ab} &=  8\pi e_a(\phi) e_b(\phi) && \text{ for the Einstein--scalar
field system} \\
M_{ab} &= 16\pi\mu u_a u_b && \text{for the Einstein--stiff fluid system}
\end{align}
\end{subequations}
$M^a_{\ b}$ will be estimated in section \ref{sec:curvest}. Note that the
power of $t$ occurring on the right hand side of equation (\ref{einsteinnsg})
is positive due to (\ref{alphacombo}).

The wave equation (\ref{scalar}) becomes the following system of 
equations for $\psi,\omega_a,\chi$.
\begin{subequations}\label{eq:wave-first}
\begin{align}
t\d_t \psi + \beta \psi -\chi &= 0 
 \\
t\d_t\omega_a&=t^\zeta[e_a(\chi)+(\zeta-\beta)e_a(\psi)]    \\     
t\d_t \chi + \beta \chi  &= 
t^{\alpha_0 - \beta} \tr \kappa (A + t^{\beta} \chi) 
+ t^{2-\beta} \Delta \ophi + t^{2-\zeta} \nabla^a \omega_a
\end{align}
\end{subequations}

Let $\calU = (\gamma^a_{\ b}, \kappa^a_{\ b}, \psi,  \chi, 
\lambda^a_{bc},\omega_a)$. Then we can write the second 
reduced system in the Einstein--scalar field case, which consists
of equations (\ref{eq:ein-first-red}) and (\ref{eq:wave-first}) in the 
form 
$$
t\d_t \calU + \calA \calU = \calF(t,x,\calU,\calU_x) .
$$
for a matrix $\calA$ and a function $\calF$. 
We will prove that this system is in Fuchsian form.

The Einstein-stiff fluid equations will be treated in a similar way. However,
due to the complexity of the equations in that case, they will not be written
out more explicitly than is absolutely necessary to understand the essential
features of their structure. The second reduced system in the stiff fluid case
can be brought into the generalized Fuchsian form 
\begin{equation}\label{genfuchs}
t\d_t \calU + \calA \calU = \calF(t,x,\calU,\calU_x)
+\calG(t,x,\calU)\d_t \calU 
\end{equation}
already introduced in section \ref{sec:fuchs}. In order to do this it is 
useful to introduce some abbreviations for certain terms in (\ref{euler}) 
so that the equations become
\begin{subequations}
\begin{align}
&&\d_t\mu-2(\tr k)\mu=-2|u|^2\d_t\mu-4\mu u^a\d_t u_a
+F_1               \\
&&\d_t u_a+(\tr k)u_a+(1/2)\mu^{-1}e_a(\mu)
=-\mu^{-1}[\d_t \mu-2(\tr k)\mu]u_a                           \\
&&+(1+|u|^2)^{-1}u^b u_a\d_t u_b\nonumber                       
-[(1+|u|^2)^{-1/2}-1]\mu^{-1}e_a(\mu)+F_2
\end{align}
\end{subequations}
The expressions $F_1$ and $F_2$ contain only terms which can be incorporated 
into $\calF$ in (\ref{genfuchs}). Next the ansatz (\ref{eq:stiffansatz})
must be substituted into these equations. The result is:
\begin{subequations}\label{eq:nuu}
\begin{align}
t\d_t\nu+\beta_1\nu&=-2t^{3-\beta_1}(1+t\tr k){}^0\mu+2(1+t\tr k)\nu
+t^{3-\beta_1}[\d_t \mu-2(\tr k)\mu]    \\
t\d_t v_a+\beta_2 v_a&=-(1+t\tr k)v_a+t^{-\beta_2}[\d_t u_a+(\tr k)u_a
+(1/2){}^0\mu^{-1}e_a({}^0\mu)]
\end{align}
\end{subequations}
The expressions on the left hand side of the above form of the Euler 
equations written in terms of the basic variables $\mu$ and $u_a$ occur
on the right hand sides of the above evolution equations for $\nu$ and $v_a$.
In order to get a fully explicit form it would be necessary to substitute for
these expressions and then express the final result in terms of $\nu$ and
$v_a$. This is, however, neither necessary nor even helpful for the analysis
to be done here. 

Next we will consider the matrix $\calA$ and prove that $\calA$ is a direct 
sum of a matrix with spectrum bounded from below by a positive number, 
with a zero matrix. (The arguments in the scalar field and stiff fluid cases 
are very similar.) It is therefore of a form such that the theory presented
in section \ref{sec:fuchs} applies. In addition we must show that 
$\calF(t,x,\calU, \calU_x) = O(t^{\delta})$ for some $\delta > 0$ and, in
the stiff fluid case, that $\calG(t,x,\calU)$ satisfies a similar estimate.
This will be done in the next section. 

The matrix $\calA$ is block diagonal and therefore it is
enough to consider each block separately. The rows and columns of $\calA$
corresponding to $\lambda^a_{\ bc}, \omega_a$ are zero, and therefore this
$\calA$ is the direct sum of a matrix corresponding to $\gamma, \kappa,
\psi,\chi$, with a zero matrix in the scalar field case and the direct sum 
of a matrix corresponding to $\gamma, \kappa, \nu, v_a$ with a zero matrix
in the stiff fluid case. We now consider the spectrum of this matrix. 
The submatrix corresponding to $\gamma,\kappa$ 
is upper block triangular. 
The $\gamma,\gamma$ block is given by
$$
\gamma^a_{\ b} \mapsto \alpha^a_{\ b}\gamma^a_{\ b} + 2 [\gamma , t\ok]^a_{\ b}
$$
To estimate the spectrum of this, it is necessary to consider the cases
I,II,III separately. 
Working in a frame which diagonalizes $\ok$, $t\ok^a_{\ b} = - \delta^a_{\
b} p_b$, and hence in this case 
$$
2 [ \gamma , t\ok]^a_{\ b} = -2(p_b - p_a) \gamma^a_{\ b}
$$
Therefore, in case III, we get using the definition of $\alpha^a_{\ b}$,  
$$
\alpha^a_{\ b}\gamma^a_{\ b} + 2 [\gamma , t\ok]^a_{\ b} = 2( (p_b - p_a)_+ 
 - (p_b - p_a)+ \alpha_0 ) \gamma^a_{\ b}
$$
and hence using $(x)_+ - x \geq 0$ for all $x \in \R$, 
the spectrum of the $\gamma,\gamma$ block is bounded from below by
$\alpha_0$ in case III. 

Next consider case I. In this case, $\alpha^a_{\ b} = \alpha_0 = 4\eps$ and 
the spectrum of 
$\gamma^a_{\ b} \mapsto 2 [\gamma , t\ok]^a_{\ b}$ is bounded from
below by $\eps$, which shows that the spectrum of the $\gamma,\gamma$ block 
is bounded from below by $3\eps$ in case II. 

Finally, in case II, $\ok^a_{\ b}$ is block diagonal in the adapted
frame. The spectrum of 
$\gamma^a_{\ b} \mapsto 2 [\gamma , t\ok]^a_{\ b}$ 
consists of $2(p_{a'}-p_{b'})$, $2(p_{a'} - p_{\perp})$, and $0$. Now using
the definition of $\alpha^a_{\ b}$ for case II and arguing as above, we get
the lower bound $3\eps$ for the spectrum of the $\gamma,\gamma$ block in
case II. 
Therefore the spectrum of the $\gamma,\gamma$ of $\calA$ is bounded from
below by $3\eps$. 

Next we consider the $\kappa,\kappa$ block. This is of the form 
$$
\kappa^a_{\ b} \mapsto \alpha^a_{\ b} \kappa^a_{\ b} - (t \ok^a_{\ b} ) \tr
\kappa
$$
First consider the action on the trace--free part of $\kappa^a_{\ b}$. Then
the spectrum is given by $\alpha^a_{\ b} > \alpha_0$. On the other hand,
restricting to the trace part of $\kappa^a_{\ b}$, which is diagonal, we see
that the spectrum is $\alpha_0 +1$. Therefore the spectrum of the
$\kappa,\kappa$ block is bounded from below by $\alpha_0$. 

The $\psi,\chi$ block is of the form 
$$
\begin{pmatrix} \beta & -1 \\
                  0 & \beta 
\end{pmatrix} 
$$
which has spectrum $\beta > 0$. The $\nu, v_a$ block is diagonal with 
eigenvalues $\beta_1$ and $\beta_2$.

Therefore, in view of the facts that $3\eps > \beta > 0$, $\beta_1>0$ and 
$\beta_2>0$, the desired properties of the spectrum of $\calA$
have been verified. 

Given a solution 
$\calU = 
(\gamma^a_{\ b}, \kappa^a_{\ b}, \psi, \chi, \lambda^a_{bc}, \omega_a)$ 
of the reduced system 
for the Einstein--scalar field equations, 
define $g_{ab}$, $k_{ab}$ and $\phi$ 
by (\ref{eq:ansatz}) and (\ref{eq:ansatzp}). Similarly, given a solution 
$\calU = 
(\gamma^a_{\ b}, \kappa^a_{\ b}, \nu, v_a \lambda^a_{bc})$ of the reduced
system for the Einstein--stiff fluid equations, define $(g_{ab}, k_{ab}, \mu, 
u_a)$ by (\ref{eq:ansatz}) and (\ref{eq:stiffansatz}). 

If it can be shown that $g_{ab}$ and $k_{ab}$ are symmetric then a solution 
of the Einstein-scalar field equations is obtained.
The next lemma gives sufficient conditions for this to be true.
\begin{lemma}\label{symmetry}
Let a solution of the velocity dominated Einstein-matter system be given on 
$S\times (0,\infty)$ with all eigenvalues of $-t\ok^a{}_b$ positive and 
$t\tr \ok=-1$. Let $\calU$ be a solution of the reduced system for the 
Einstein--scalar field or Einstein--stiff fluid system corresponding to 
the given velocity dominated solution, with $\calU = \ordo(1)$. Define 
$(g_{ab},k_{ab},\phi)$ by (\ref{eq:ansatz}) and (\ref{eq:ansatzp}) in the 
scalar field case, and define $(g_{ab}, k_{ab}, \mu, u_a)$ by 
(\ref{eq:ansatz}) and (\ref{eq:stiffansatz}) in the stiff fluid case. 
Then $g_{ab}$ and $k_{ab}$ are symmetric. 
\end{lemma}
\begin{proof}
From the evolution equation for $\gamma^a_{\ b}$ and the 
definitions of $g_{ab}$ and $k_{ab}$ it follows that $\d_t g_{ab}=-2k_{ab}$. 
Similarly an equation close to the usual evolution equation for $k^a_{\ b}$ 
can be recovered from (\ref{einsteinnsk}). It differs from the usual one only 
in the fact that $R^a_{\ b}$ is replaced by $\SR^a_{\ b}$. From these 
equations we can derive the equations:
\begin{subequations}\label{antisymm}
\begin{align}
\d_t(g_{ab}-g_{ba})&=-2(k_{ab}-k_{ba})\label{antisymmg}      \\
\d_t(k_{ab}-k_{ba})&=(\tr k)(k_{ab}-k_{ba})\label{antisymmk}
\end{align}
\end{subequations}
It follows from the
assumptions on $\gamma^a_{\ b}$ and $\kappa^a_{\ b}$ together with the
definition of $k_{ab}$, that the components of
$k_{ab}-k_{ba}$ are $o(t^{-1+\eta})$ for some $\eta>0$. Hence the quantity
$\Omega_{ab}=t^{1-\eta}(k_{ab}-k_{ba})$ tends to zero as $t\to 0$.
It satisfies the equation:
\begin{equation}
t\d_t \Omega_{ab}+\eta\Omega_{ab}=(t\tr k+1)\Omega_{ab}
\end{equation}
From Theorem \ref{fuchstheorem} we conclude that $\Omega_{ab}=0$. Thus 
$k_{ab}$ is symmetric. It then follows immediately from (\ref{antisymmg}) 
and the fact that $g_{ab}=o(1)$ that $g_{ab}$ is also symmetric.
\end{proof}

\section{Curvature estimates}\label{sec:curvest}
Let $\SR_{ab}$ be the Ricci tensor computed w.r.t. the symmetrized metric
$\Sg_{ab} = \half ( g_{ab} + g_{ba})$.
In order to get a Fuchsian form for the Einstein--matter evolution
equations, we need the following estimate for the frame components of
$\SR^a_{\ b}$,  
\begin{equation}\label{eq:firstSRest}
t^{2-\alpha^a_{\ b}} \SR^a_{\ b} = O(t^{\delta}), \quad \text{ for some } 
\delta \in (0,\eps).
\end{equation}

In doing the estimates we will use the notion of comparing the size of 
functions introduced in Definition \ref{def:oleqdef}. 
In proving that the second reduced system 
$$
t \d_t u + A(x) u = f(t,x,u,u_x)
$$
is in Fuchsian form, one essential step is to prove an estimate of the form 
$$
f \oleq t^{\delta}
$$
for some $\delta > 0$. In the present section, we accomplish this task for
the expression $t^{2-\alpha^a_{\ b}} \SR^a_{\ b}$, which is now considered as 
a function $r(t,x,v(x,t),u,u_x)$, where $v(x,t)$ is defined in terms of the
solution $\og_{ab}, \ok_{ab}$ to the velocity dominated system 
and the data $\{e_a\}, \{\theta^a\}, \{q_a\}$, etc. defined in 
section \ref{sec:setup}, and $u$ consists of the variables 
$\gamma^a_{\ b}, \lambda^a_{\ bc}$. In terms of the relation $\oleq$, the 
goal is to prove 
\begin{equation}\label{eq:SRest}
t^{2-\alpha^a_{\ b}} \SR^a_{\ b} \oleq t^{\delta}, \quad \text{ for some } 
\delta \in (0,\eps).
\end{equation}
By assumption, $\alpha_0 = 4 \eps$, so 
$\alpha^a_{\ b} - 2\eps = 2(q_b - q_a)_+ + \alpha_0/2$. 
Therefore, the
arguments that apply to $\alpha^a_{\ b}$ also apply to $\alpha^a_{\ b} -
2\eps$. 
 
The symmetrized metric tensor satisfies 
\begin{subequations}\label{eq:Sansatz}
\begin{align}
\Stg_{ab} &= \tog_{ab} + t^{\talpha^a_{\ b} - 2\eps}\, \tog_{ac}
{}^S \gamma^c_{\ b} , \qquad {}^S\gamma^c_{\ b} = \ordo(1) , \\
\Stg^{ab} &= \tog^{ab} + t^{\talpha^a_{\ b} - 2 \eps} \, {}^S \bar
\gamma^a_{\ c} \tog^{cb}, \qquad {}^S \bar \gamma^a_{\ c} = \ordo(1) 
\end{align}
\end{subequations}
To see this, note the identity
$$
{}^S \gamma^c_{\ b} = \half t^{2\eps} \left ( 
\gamma^c_{\ b} + \tog^{cd}\tog_{bf} \gamma^f_{\ d} \right ) ,
$$
which in view of Lemma \ref{lem:gest} shows that ${}^S \gamma^c_{\ b} =\ordo(1)
$. The argument that ${}^S \bar \gamma^a_{\ c} = \ordo(1)$ is the
same as for $\bar \gamma^a_{\ c} =\ordo(1)$. 

It is convenient to estimate the rescaled frame components. 
Note $R^a_{\ b} = t^{-q_a + q_b} \tR^a_{\ b}$. Hence in view of
(\ref{eq:talpha}) 
we need to consider 
$$
t^{2 - \talpha^a_{\ b}} \tR^a_{\ b}
$$
Using Lemma \ref{lem:gest} and (\ref{eq:ansatz}) gives 
$$
|| t^{2 - \talpha^a_{\ b}} \tR^a_{\ b}||_\infty \leq 
C t^{-\eps} ||t^{2-\talpha^a_{\ b}} \tR_{ab} ||_\infty
$$
To see this, we compute using (\ref{eq:galpha}) and (\ref{eq:tgup})
\begin{align*}
||t^{2 - \talpha^a_{\ b}} \tR^a_{\ b}||_\infty &= 
||t^{2 - \talpha^a_{\ b}} \tg^{ac} \tR_{cb} ||_\infty \\
&\leq  ||t^{2 - \talpha^a_{\ b}} \tog^{ac} \tR_{cb}||_\infty + 
||t^{2 - \talpha^a_{\ b}+ \talpha^a_{\ c} } 
  \bar \gamma^a_{\ d} \tog^{dc} \tR_{cb} ||_\infty \\ 
&\leq C || t^{2 - \eps - \talpha^c_{\ b}} \tR_{cb}||_\infty
\end{align*}
where we used the triangle inequality in the form
$- \talpha^a_{\ b} + \talpha^a_{\ c} \geq
- \talpha^c_{\ b} $.

Let $\gamma^c_{ab} = \theta^c ( [e_a , e_b])$ be the structure coefficients of
the frame $\{e_a\}$.
The structure coefficients $\tComm_{ab}^c= \ttheta^c ( [ \te_a , \te_b ] )  $ 
of the frame $\te_a$ are given by
\begin{equation}\label{eq:tstruct}
\tComm_{ab}^c = 
t^{q_c - q_a -q_b} \Comm_{ab}^c
- \log(t) ( t^{-q_a} e_a (q_b)\delta^c_b 
- t^{-q_b} e_b (q_a)\delta^c_a  )
\end{equation}

It is convenient to define 
$
\tGamma_{abc} = \la \nabla_{\te_a} \te_b , \te_c \ra .
$
Then $\tGamma_{abc}$ is given in terms of $\tg_{ab}$ by
\begin{equation}\label{eq:tGamma}
\begin{split}
2 \tGamma_{abc} &= 
\te_a(\tg_{bc}) + \te_b(\tg_{ac})
- \te_c(\tg_{ab}) \\
&\quad
+ \tComm_{ab}^d \tg_{cd} - \tComm_{ac}^d \tg_{bd} - \tComm_{bc}^d \tg_{ad}
\end{split}
\end{equation}
and $\tR_{dcab}$ is given in terms of $\tGamma_{abc}$ and $\tComm^a_{\ bc}$ 
by   
\begin{align}
\tR_{dcab} &= \te_a \tGamma_{bcd}
- \te_b \tGamma_{acd} 
- \tComm^f_{\ ab} \tGamma_{fcd} \nonumber \\
&\quad
- \tg^{fg} \tGamma_{bcf} \tGamma_{adg} 
+ \tg^{fg} \tGamma_{acf} \tGamma_{bdg} \nonumber \\
&= (R1)_{dcab} - (R2)_{dcab} - (R3)_{dcab} - (R4)_{dcab} + (R5)_{dcab} 
\label{eq:Rdef}
\end{align}

Let 
$$
z_{ab} 
= \left \{ \begin{array}{cl} 
0, & \text{ if } a=b \\ 1, & \text{ if } a \ne b \end{array} \right.
$$

Define 
\begin{equation}\label{eq:Zdef}
\begin{split}
Z_{abc}(t) &= t^{-q_a} + t^{-q_b} + t^{-q_c} \\
&\quad  
+ t^{q_a - q_b - q_c} z_{bc} + t^{q_b - q_c - q_a} z_{ca} 
+ t^{q_c - q_a - q_b} z_{ab} 
\end{split}
\end{equation}
Note that $(R1)_{dcab},\dots, (R5)_{dcab}, z_{ab}, Z_{abc}$ are not tensors.
In the rest of this section, the
frame $\{e_a\}$ is fixed and the estimates will be done for tensor
components in this frame.

We will use the following lemma as the starting point of the estimates in 
this section. 
\begin{lemma}\label{lem:goleq}
\begin{subequations}\label{eq:goleq}
\begin{align}
\tg_{ab} &\oleq t^{-\eps} \label{eq:geps}\\
\tg^{ab} &\oleq t^{-\eps} \label{eq:ginveps}\\
\te_a \tg_{bc} &\oleq t^{-q_a - 2\eps} \label{eq:g1eps}\\
\te_a \te_b \tg_{cd} &\oleq t^{-q_a - q_b - 3\eps} \label{eq:g2eps} \\
\tg^{ab} t^{-q_b} &\oleq t^{-\eps - q_a} & 
\label{eq:tgabqb} \\
\te_c (t^{q_a} \tg_{ab} ) &\oleq t^{-q_c + q_b - 2\eps} \label{eq:gswitch} \\
\tg^{ab}Z_{abc} &\oleq t^{-\eps} (t^{-q_h} + t^{-q_c} + t^{q_c - 2q_h})
& h = \min(a,b) \label{eq:tgabZabc} \\
\tComm^c_{ab} \tg^{bd} &\oleq t^{-2\eps} (t^{q_c - q_a - q_d} + t^{-q_a} +
t^{-q_d} ) & \label{eq:tCommtg} 
\end{align}
\end{subequations}
\end{lemma}
\begin{proof}
The inequalities (\ref{eq:geps}) and (\ref{eq:ginveps}) are
immediate from Lemma \ref{lem:gest}, 
(\ref{eq:tansatz}) and definition \ref{def:oleqdef}. 
Recalling that the variables $u,u_x$ occurring in the second reduced system
contain $\gamma^a_{\ b}$, $\lambda^a_{bc}$ and its first order derivatives 
gives (\ref{eq:g1eps}) and (\ref{eq:g2eps}). (Here the inequality 
$\zeta<\epsilon$ has been used.)  

The estimate (\ref{eq:tgabqb}) follows
from Lemma \ref{lem:gest}, (\ref{eq:VDrel}) and (\ref{eq:tansatz}) together
with the triangle inequality in the form $-q_b \leq -q_a + |q_a -
q_b|$. The estimate (\ref{eq:gswitch}) follows from
(\ref{eq:VDrel}) and (\ref{eq:tansatz}) together with the observation that
since $\og_{ab}$ is block diagonal for $x \in U_0$, 
$e_c \og_{ab}$ is also block diagonal.
The estimates (\ref{eq:tgabZabc}) and 
(\ref{eq:tCommtg}) follow in a similar way starting from
(\ref{eq:Zdef}) and (\ref{eq:tstruct}).
In (\ref{eq:tCommtg}) a $\log(t)$ term is dominated by $t^{-\eps}$. 
\end{proof}

The following lemma gives estimates of $\tGamma_{abc}$ in terms of
$Z_{abc}$
\begin{lemma}\label{lem:Gamest}
\begin{subequations}\label{eq:Gamest}
\begin{align}
\tGamma_{abc} &\oleq t^{-2\eps} Z_{abc} \label{eq:GamZ} \\
\tGamma_{abb} &\oleq t^{-2\eps - q_a} \label{eq:GamZabb} \\
\tGamma_{aab} &\oleq t^{-2\eps} (t^{-q_a} + t^{-q_b}) \\
\tGamma_{aba} &\oleq t^{-2\eps} (t^{-q_a} + t^{-q_b}) \\
\te_a \tGamma_{bcd} &\oleq t^{-3\eps - q_a} Z_{bcd} \label{eq:tetGamma}
\end{align}
\end{subequations}
\end{lemma}
\begin{proof}
First observe that 
$\Comm^f_{bc} \oleq z_{bc}$. From (\ref{eq:tstruct}) and (\ref{eq:goleq})
we have 
$$
\tComm^f_{bc} \tg_{fa} \oleq t^{-2\eps} ( t^{q_a - q_b - q_c}z_{bc} + t^{-q_b} 
+ t^{-q_c} ) 
$$
Now noting $\te_a(\tg_{bc}) \oleq t^{-q_a - 2\eps}$, (\ref{eq:GamZ}) follows.

To estimate $\tGamma_{abb}$ we compute 
$$
\tGamma_{abb} = \half t^{-q_a} e_a \tg_{bb} \oleq t^{-2\eps - q_a}
$$
The estimates for $\tGamma_{aab}, \tGamma_{aba}$ follow directly from
(\ref{eq:GamZ}) and the definition of $Z_{abc}$. 

Finally we consider (\ref{eq:tetGamma}). Expanding out $\te_a \tGamma_{bcd}$, 
we see that it contains (up to permutations of the indices) 
terms of the form $\te_a\te_b (\tg_{cd})$, 
$(\te_a \tComm^f_{cd})\tg_{fb}$ and $\tComm^f_{cd} \te_a \tg_{fb}$. The first 
and second type of terms are estimated using (\ref{eq:g2eps}) and
(\ref{eq:tgabqb}), using the form of $\tComm^f_{cd}$. Finally, the third
type of term is estimated using (\ref{eq:gswitch}).
\end{proof}

An important consequence of the Kasner relation $\sum_a q_a = 1$, is 
\begin{equation}\label{eq:q1-rel}
2 + 2(q_1 - q_2 - q_3) = 4q_1
\end{equation}
which implies 
\begin{equation}\label{eq:q1-oleq}
t^{2+2(q_j - q_k - q_l)} \oleq t^{4q_1} \qquad \text{\em if at least one of 
$k,l$ is different from  3},
\end{equation} 
The strategy will be to eliminate as much as possible the occurence of
repeated negative exponents, in order to be able to use this relation. 


We make note of the following useful relations.
\begin{subequations}\label{eq:Zrel-a}
\begin{align}
Z_{abc} &\oleq t^{q_1 - q_2 - q_3} & \label{eq:Z123} \\
Z_{hhc} &\oleq t^{-q_h} + t^{-q_c} & \label{eq:Zhhc} \\
e_a Z_{bcd} &\oleq t^{-\eps} Z_{bcd} & \label{eq:derZ-a}
\end{align}
\end{subequations}
The estimates 
(\ref{eq:Z123}) and (\ref{eq:Zhhc}) are immediate from (\ref{eq:Zdef}).

For the rest of this section, we will assume that $g_{ab}$ is symmetric and
is of the form given by  (\ref{eq:gdef}). Let $R_{ab}$ be the Ricci tensor
defined with respect to $g_{ab}$. Under these assumptions we will estimate 
$R^a_{\ b}$. The estimate then applies after a small
modification to $\SR^a_{\ b}$.

We now proceed to estimate the rescaled components 
of the Ricci tensor $\tR_{ad} = \tg^{bc} \tR_{dcab}$. Corresponding to the
terms $(R1),\dots, (R5)$ we have
$$
\tR_{ad} = (\Ric1)_{ad} - (\Ric2)_{ad} - (\Ric3)_{ad} - (\Ric4)_{ad} + (\Ric5)_{ad}
$$
We make the following simplifying observations.
\begin{itemize}
\item By the symmetry $\tR_{ad} = \tR_{da}$ we can assume without loss of
generality that $a\leq d$. 
\item $\tR_{dcab}$ is skew symmetric in the first and second pair of
indices, therefore we can assume without loss of generality that
$c\ne d$, $b \ne a$. 
\end{itemize}

Therefore, in the following, we can without loss of generality use the 
following {\bf convention}:
The indices $a,b,c,d$ satisfy the relations
\begin{equation}\label{eq:cond}
a\leq d, \qquad c \ne d, \qquad a \ne b
\end{equation}

We will now estimate $\tR_{ad}$ by considering each term $(\Ric1)_{ad},
\dots, (\Ric5)_{ad}$ in turn. 
The estimate we will actually prove is of the form 
$$
t^{2 + q_a - q_d} \tR_{ad} \oleq t^{4q_1 - 6 \eps}
$$ 
which 
will imply the needed estimate for $\SR^a_{\ d}$. 

\subsection{$(\Ric1)$}
$$
(\Ric1)_{ad} = \tg^{bc} \te_a \tGamma_{bcd}
$$ 
where we are summing over repeated indices. 
Let $F = t^{2+q_a - q_d} (\Ric1)_{ad}$. 

Using Lemma \ref{lem:Gamest} and (\ref{eq:tgabZabc}), 
we have 
\begin{align*}
F &\oleq t^{2-4\eps} t^{q_a - q_d} t^{-q_a} (t^{-q_c} + t^{-q_d} + t^{q_d -
2q_c} ) \\
&\oleq t^{2-4\eps} (t^{-q_d - q_c} + t^{-2q_d} + t^{-2q_c} ) \\
&\oleq t^{4q_1 - 4\eps} \oleq t^{4q_1 - 6 \eps} 
\end{align*}

\subsection{$(\Ric2)$}
$$
(\Ric2)_{ad} = \tg^{bc} \te_b \tGamma_{acd}
$$
Let $F = t^{2+q_a - q_d} (\Ric2)_{ad}$. By Lemma \ref{lem:Gamest} and 
(\ref{eq:tgabqb}),
$$
F \oleq t^{2-4\eps} t^{q_a - q_d - q_c} Z_{acd}
$$
By (\ref{eq:cond}), $d \ne c$ and using (\ref{eq:Z123}) and 
(\ref{eq:q1-oleq}) this gives 
$$
F \oleq t^{4q_1 - 4 \eps} \oleq t^{4q_1 - 6 \eps}
$$
which is the required estimate.

\subsection{$(\Ric3)$}
$$
(\Ric3)_{ad} = \tg^{bc} \tComm^f_{ab} \tGamma_{fcd}
$$
Let $F = t^{2+q_a - q_d} (\Ric3)_{ad}$. 
We estimate using (\ref{eq:tCommtg}) and Lemma \ref{lem:Gamest},
\begin{align*}
F &\oleq t^{2-4\eps} t^{q_a - q_d} (t^{q_f-q_a - q_c} + t^{-q_a} + t^{-q_c}
) Z_{fcd} \\
&\oleq t^{2-4\eps} (t^{q_f - q_d - q_c} + t^{-q_d} + t^{q_a - q_d - q_c})
Z_{fcd} \\
\intertext{use $c \ne d$ by (\ref{eq:cond}), (\ref{eq:Z123}) and
(\ref{eq:q1-oleq})}
&\oleq t^{4q_1 - 4\eps} \oleq t^{4q_1 - 6 \eps}
\end{align*}

\subsection{$(\Ric4)$}
$$
(\Ric4)_{ad} = \tg^{bc} \tg^{fg} \tGamma_{bcf} \tGamma_{adg} 
$$
Let $F = t^{2+q_a - q_d} (\Ric4)_{ad}$.
We have by Lemma \ref{lem:Gamest},
$$
F \oleq t^{2+q_a - q_d - 4\eps}  \tg^{bc} \tg^{fg} 
Z_{bcf} Z_{adg} 
$$
In case $a=d$ this gives, with $h=\min(b,c)$, $m \in \{f,g\}$, using
(\ref{eq:Zrel-a}) and (\ref{eq:goleq}), 
\begin{align*}
t^2 (\Ric4)_{aa} &\oleq t^{2-6\eps} (t^{-q_h} + t^{-q_m} + t^{q_m - 2q_h} ) 
(t^{-q_a} + t^{-q_m} ) \\
&\oleq t^{4q_1 - 6 \eps} 
\end{align*}
Next we consider the case $a < d$. In case $g=d$, Lemma \ref{lem:Gamest}
together with (\ref{eq:tgabqb}) and (\ref{eq:tgabZabc}) gives
\begin{align*}
F &\oleq t^{2+q_a - q_d} \tg^{bc} \tg^{fd} \tGamma_{bcf} t^{-2\eps - q_a} \\
&\oleq t^{2-6\eps} ( t^{-q_f -q_h} + t^{-2q_f} + t^{-2q_h}) \qquad h = \min(b,c) \\
&\oleq t^{4q_1 - 6 \eps} 
\end{align*}
In case $a=g$ we have arguing as above
\begin{align*}
F &\oleq t^{2+q_a - q_d} \tg^{bc} \tg^{fa} \tGamma_{bcf} 
t^{-2\eps} (t^{-q_a} + t^{-q_d} ) \\
&\oleq t^{2-5\eps} \tg^{fa} (t^{-q_h} + t^{-q_f} + t^{q_f - 2q_h})
(t^{-q_d} + t^{q_a - 2 q_d} ) , \qquad h = \min(b,c) \\
&\oleq t^{2-6\eps} (t^{-q_h} + t^{-q_m} + t^{q_m - 2q_h})
(t^{-q_d} + t^{q_m - 2 q_d} ), \qquad m = \min(a,f)
\end{align*}
From $h=\min(b,c)$ and $c\ne d$ which holds by (\ref{eq:cond}), we find that 
either $h<3$ or $d<3$ must hold. Using this it follows using 
(\ref{eq:q1-oleq}) that in case $a=g < d$, 
$F \oleq t^{4q_1 - 6 \eps}$.

It remains to consider the case when $a<d$ and $a,d,g$ are distinct.
In this case, the estimates used above give
\begin{align*}
F 
&\oleq t^{2+q_a - q_d - 6\eps} ( t^{-q_h} + t^{-q_g} + t^{q_g - 2q_h} ) 
Z_{adg} , \qquad h=\min(b,c) \\
&\oleq t^{2- 6\eps} ( t^{-q_h} + t^{-q_g} + t^{q_g - 2q_h} )  \\
&\quad 
( t^{-q_d} + t^{q_a - 2q_d} + t^{q_a - q_d - q_g} 
+ t^{2q_a - 2q_d - q_g} + t^{-q_g} +
t^{q_g-2q_d} ) 
\end{align*}
By construction, $h < 3$ or $d < 3$ must hold, which in conjunction with the
fact that in the present case, $g \ne d$ gives using (\ref{eq:q1-oleq}), 
$$
F \oleq t^{4q_1 - 6 \eps} 
$$
The above proves that the required estimate 
$t^{2+q_a - q_d} (\Ric4)_{ad} \oleq t^{4q_1 - 6\eps}$ holds. 

\subsection{$(\Ric5)$}
$$
(\Ric5)_{ad} = \tg^{bc} \tg^{fg} \tGamma_{acf} \tGamma_{bdg}
$$
The estimate for $(\Ric5)_{ad}$ is the most complicated, and will be done in
several steps. We review the steps which will be used here. In each step the
conditions on the indices $a,c,f,b,d,g$ which leads to the required estimate
may be excluded from our considerations. Recall that $a \leq d$ may be
assumed and also note by (\ref{eq:cond}) we may
assume without loss of generality that $a\ne b$, $c\ne d$. 

The steps we will use are:
\begin{enumerate}
\item \label{point:a=d} $a=d$ can be excluded, so $ a < d$ may be assumed. 
\item \label{point:g=d} $g=d$ can be excluded, so $g\ne d$ may be assumed. 
\item \label{point:g=a} $g=a$ can be excluded, so $g\ne a$ may be assumed. 
\item \label{point:b=d} $b=d$ can be excluded, so $b\ne d$ may be assumed. 
\item \label{point:a=c} $a=c$ can be excluded, so $a\ne c$ may be assumed. 
\end{enumerate}
When all the above claims are verified, we may restrict our considerations
to the indices satisfying the conditions 
\begin{equation}\label{eq:indcond}
a<d, \quad a\ne b, \quad c \ne d, \quad g\ne d, \quad g\ne a, \quad b\ne d,
\quad a \ne c
\end{equation}
These conditions imply that the indices $\{a,d,g\}$, $\{a,d,c\}$ and
$\{a,d,b\}$ are distinct, so (\ref{eq:indcond}) implies $g=b=c$, 
as all indices take values in $\{1,2,3\}$. 
Therefore the required estimate
for $(\Ric5)_{ad}$ will hold if we can verify that it holds under
(\ref{eq:indcond}) in conjunction with the condition $g=b=c$, which is the
final step.  

Let 
$$
F = t^{2+q_a-q_d} (\Ric5)_{ad} . 
$$
\bigskip

\noindent{\bf Case $a=d$:}
In case $a=d$, Lemma \ref{lem:Gamest} and (\ref{eq:Z123}) give  
\begin{align*}
F &\oleq t^{2-6\eps} t^{2(q_1 - q_2 - q_3)} \\
&\oleq t^{4q_1 - 6 \eps} 
\end{align*} 
Therefore we may assume $a< d$ in the following. Further by (\ref{eq:cond}),
$a\ne b$ and $c\ne d$. 

\noindent{\bf Case $g=d$:}
Next consider the case $g=d$. Then using Lemma \ref{lem:Gamest},
(\ref{eq:Zdef})  and
(\ref{eq:tgabqb}) we have 
\begin{align*}
F &= t^{2+q_a - q_d} \tg^{bc} \tg^{fd} \tGamma_{acf} \tGamma_{bdd} \\
&\oleq  t^{2+q_a-q_d} \tg^{bc} \tg^{fd} \tGamma_{acf} t^{-2\eps - q_b} \\
&\oleq t^{2 - 6 \eps + q_a - q_d - q_c} (t^{-q_a} + t^{-q_c} + t^{-q_d}
+ t^{q_a - q_c - q_d} + t^{q_c - q_d - q_a} + t^{q_d - q_a - q_c} ) \\
&\oleq t^{2-6\eps} ( t^{-q_d - q_c} + t^{q_a - q_d - 2 q_c} \\
&\quad 
+ t^{q_a - 2q_d - q_c} + t^{2(q_a - q_d - q_c )} + t^{-2q_d} + t^{-2q_c}) .
\end{align*} 
By (\ref{eq:cond}),  $d \ne c$, this gives 
$F \oleq t^{4q_1-6 \eps}$ in the case $g=d$. 
\bigskip

\noindent{\bf Case $g=a$:}
Next consider the case $g=a$. Then we have 
\begin{align*}
F &= t^{2+q_a - q_d} \tg^{bc} \tg^{fa} \tGamma_{acf} \tGamma_{bda} \\
\intertext{use Lemma \ref{lem:Gamest} and (\ref{eq:tgabZabc})}
&\oleq t^{2-5\eps}t^{q_a - q_d} \tg^{bc} 
( t^{-q_a} + t^{-q_c} + t^{q_c - 2q_a} ) Z_{bda} \\
\intertext{use Lemma \ref{lem:gest}}
&\oleq t^{2-6\eps} ( t^{-q_d} + t^{q_a - q_d - q_c} + t^{q_c - q_a - q_d} )
Z_{bda} \\
\intertext{use (\ref{eq:Z123}) and $a < d$, $d\ne c$}
&\oleq t^{4q_1 - 6 \eps} 
\end{align*}

At this stage we may assume 
\begin{equation}\label{eq:tempcond}
a<d, \quad a\ne b, \quad c \ne d, \quad g \ne d, \quad g \ne a
\end{equation}
\bigskip

\noindent{\bf Case $b=d$:}
Next consider the case $b=d$. In this case
\begin{align*}
F &= t^{2+q_a - q_d} \tg^{dc} \tg^{fg} \tGamma_{acf} \tGamma_{ddg} \\
&\oleq t^{2+q_a - q_d} \tg^{dc} \tg^{fg} \tGamma_{acf} t^{-2\eps} (
t^{-q_d} + t^{-q_g} ) \\
&\oleq t^{2-6\eps} ( t^{q_a - q_d - q_c} + t^{q_a - q_d - q_g} ) Z_{acf} 
\\
\intertext{use $d\ne c$ and $d \ne g$ from (\ref{eq:tempcond})}
&\oleq t^{4q_1 - 6 \eps} 
\end{align*}
\bigskip

\noindent{\bf Case $a=c$:}
Next consider the case $a=c$. In this case we have using Lemma
\ref{lem:Gamest} 
\begin{align*}
F &\oleq t^{2+q_a - q_d} \tg^{ba} \tg^{fg} t^{-2\eps} 
( t^{-q_a} + t^{-q_f} ) t^{-2\eps} Z_{bdg} \\
\intertext{use (\ref{eq:tgabqb})}
&\oleq t^{2-6\eps} (t^{-q_d} + t^{q_a - q_d - q_g} )Z_{bdg} \\
\intertext{use $g\ne d$ from (\ref{eq:tempcond}) and (\ref{eq:Z123})}
&\oleq t^{4q_1 - 6 \eps}
\end{align*}

At this stage we may assume 
\begin{equation}\label{eq:tempcond2}
a<d, \quad a\ne b, \quad c \ne d, \quad g \ne d, \quad g \ne a, \quad b \ne d, \quad a \ne c
\end{equation}
As discussed above, 
%
if we can prove that the required estimate holds under the condition 
$g=c=b$ we are done. 
\bigskip

\noindent{\bf Case $g=b=c$:} Next consider the case $g=b=c$. In this case,
after making the substitutions $g=c$ and $b=c$, 
\begin{align}
F &= t^{2+q_a -q_d} \tg^{cc} \tg^{fc} \tGamma_{acf} \tGamma_{cdc} \nonumber \\
\intertext{use Lemma \ref{lem:Gamest}}
&\oleq t^{2-3\eps} t^{q_a - q_d} \tg^{fc} \tGamma_{acf} (t^{-q_d}+ t^{-q_c})
\label{eq:F:g=b=c}
\end{align}
To estimate $F$ we must now consider the cases $f=d$, $f=c$, $f=a$
separately.
\bigskip

\noindent{\bf Case $g=b=c$ and $f=d$:}
In case $g=b=c$ and $f=d$,
we have from (\ref{eq:F:g=b=c})
\begin{align*}
F &\oleq  t^{2-3\eps} t^{q_a - q_d} \tg^{dc} \tGamma_{acd} (t^{-q_d}+
t^{-q_c}) \\
\intertext{use (\ref{eq:tgabqb}) and Lemma \ref{lem:Gamest}}
&\oleq t^{2-6\eps} t^{q_a-q_d - q_c} Z_{acd} \\
\intertext{use $c \ne d$ from (\ref{eq:tempcond2}) and (\ref{eq:Z123})} 
&\oleq t^{4q_1 - 6 \eps}
\end{align*}
Therefore we may exclude  condition (\ref{eq:tempcond2}) in conjunction with 
$f=d$ from our considerations.
\bigskip

\noindent{\bf Case $g=b=c$ and $f=c$:}
In case $g=b=c$ and $f=c$ we have from (\ref{eq:F:g=b=c})
\begin{align*}
F&\oleq t^{2-3\eps} t^{q_a - q_d} \tg^{cc} \tGamma_{acc} (t^{-q_d}+
t^{-q_c}) \\
\intertext{use Lemma \ref{lem:Gamest}}
&\oleq t^{2-4\eps} t^{q_a - q_d} t^{-2\eps - q_a} (t^{-q_d}+
t^{-q_c}) \\
 &= t^{2-6\eps} (t^{-2q_d} + t^{-q_d - q_c}) \\
\oleq t^{4q_1 - 6 \eps}
\end{align*}
\bigskip

\noindent{\bf Case $g=b=c$ and $f=a$:}
The only remaining case is $g=b=c$ and $f=a$. In this
case we get from  (\ref{eq:F:g=b=c}) using Lemma \ref{lem:Gamest}
\begin{align*}
F&\oleq  t^{2-5\eps} t^{q_a - q_d} \tg^{ac} Z_{aca} (t^{-q_d}+
t^{-q_c}) \\
\intertext{use (\ref{eq:Zhhc}) and (\ref{eq:tgabZabc})}
&\oleq t^{2-6\eps} (t^{q_a - 2q_d} + t^{q_a - q_d - q_c}) t^{-q_a} \\
&\oleq t^{2-6\eps} (t^{-2q_d} + t^{-q_d - q_c} )
\\
&\oleq t^{4q_1 - 6\eps}
\end{align*}

Therefore it now follows that under (\ref{eq:tempcond2}),
the required estimate 
$$
F \oleq t^{4q_1 - 6\eps}
$$
holds and hence by the above argument it follows that this estimate holds
under (\ref{eq:cond}).

This proves for a symmetric metric satisfying (\ref{eq:ansatz}) 
the estimate 
\begin{equation}\label{eq:1curvest}
t^{2 - \alpha^a_{\ b}} R^a_{\ b}  \oleq t^{4q_1 - 6\eps}
\end{equation}
We wish to apply this to the symmetrized metric $\Sg_{ab}$,
which has the property that the rescaled symmetrized metric $\Stg_{ab}$ 
satisfies (\ref{eq:Sansatz}). The estimate (\ref{eq:1curvest}) translates
to an estimate for a metric satisfying (\ref{eq:Sansatz}) after replacing
$\alpha_0$ by $\alpha_0 - 2\eps$, which by the definition of $\alpha_0$
satisfies $\alpha_0 - 2\eps > \eps > 0$. Therefore we get 
in view of the discussion at the beginning of this section 
$$
t^{2-\talpha^a_{\ d}} \StR^a_{\ d} \oleq  t^{4p_1 - 9 \eps - \alpha_0}
$$
or 
$$
t^{2-\alpha^a_{\ d}} \SR^a_{\ d}  \oleq  t^{4p_1 - 9 \eps - \alpha_0}
$$
Now recall $\eps = \alpha_0/4 = \min_a \{p_a(x_0)\}/40$  and $q_1 > 20\eps$ 
by construction. 
This gives 
\begin{align*}
t^{2-\alpha^a_{\ d}} \SR^a_{\ d}  &\oleq  t^{4q_1 - 13 \eps} \\
&\oleq 
t^{3q_1} \\
&\oleq  t^{3p_1} 
\end{align*}
where we used that fact that $q_1 \geq p_1$ by construction.

This finishes the proof of 
\begin{lemma}[Curvature estimate]\label{lem:curvest}
$$
t^{2-\alpha^a_{\ b}} \SR^a_{\ b}  \oleq  t^{3p_1}
$$
\qed
\end{lemma}

Now some estimates will be obtained for matter variables. These will be
used to check that the right hand side of the second reduced system has
the properties required for a Fuchsian system.  Let $w_a$ be a one-form 
with the property that $w_a\oleq t^{q_a}$. We have 
\begin{align*} 
t^2 \tg^{ab}\nabla_a w_b 
&= t^2 \tg^{ab} \te_a w_a - t^2 \tg^{ab} \tg^{fg} \tGamma_{abf}  w_g
\\
&\oleq t^{2-\eps - 2q_a} + t^{2-4\eps}t^{q_1+q_2-q_3}t^{-q_g} \\
&\oleq t^{4q_1 - 4\eps} \oleq t^{4p_1-4\eps}
\end{align*}
Here it has been assumed that $w_a$ behaves in a suitable way upon taking
derivatives. In the context of the matter variables this will be the case
for the relevant choices of $w_a$, namely $e_a(\phi)$ and $t^{-1}u_a$.
Note that this estimate requires no use of cancellations, since it
only uses the relation (\ref{eq:Z123}) and not (\ref{eq:tgabZabc}). In this
situation the estimate for a given quantity is never more difficult than
that for the corresponding velocity dominated part, since the difference
between the two is always of higher order. This gives the estimates required 
for the matter equations in the scalar field case. For the stiff fluid some 
more work is needed.

The aim now is to estimate the terms on the right hand side of equations
(\ref{eq:nuu}) by a positive power of $t$. For most of these terms no
cancellations are required to get the desired estimate. There are only
two exceptions to this and they will be discussed explicitly now. The first 
is the following combination
which arises if the evolution equation for $v_a$ is written out explicitly: 
\begin{equation}
{}^0\mu^{-1}e_a({}^0\mu)-\mu^{-1}e_a(\mu)
\end{equation}
This expression is equal to
\begin{equation}
t^{\beta_1}[-A^{-1}\nu (1+A^{-1}\nu t^{\beta_1})^{-1}A^{-1}(\nabla_a A
+\nabla_a\nu)+A^{-1}t^{\beta_1}\nabla_a\nu]
\end{equation}
which is $O(t^{\beta_1})$. The contribution of this expression to the
right hand side of the evolution equation for $v_a$ is as a consequence
$O(t^{\beta_1-\beta_2})$ which shows that it is necessary to choose
$\beta_2<\beta_1$. The second expression where a cancellation is necessary
is $1+t\tr k$. Now $\tr k=-t^{-1}+\tr\kappa t^{-1+\alpha_0}$ and hence
$1+t\tr k=\tr\kappa t^{\alpha_0}$. It follows that this expression is
$O(t^{\alpha_0})$.

The analysis of the other terms is rather straightforward, although lengthy, 
and will not be carried out explicitly here. However some comments may be 
useful. The terms which are a priori most difficult to estimate are those 
involving covariant 
derivatives of $u_a$. For those it is convenient to use the components
in the rescaled frame $\tilde e_a$. For all other terms the original frame 
$e_a$ can be used straightforwardly. In order that all terms can be 
estimated by a positive power of $t$ it suffices to choose $\beta_1$ and
$\beta_2$ small enough. One possible choice is 
$\beta_2<\beta_1<q_1-5\epsilon$.

In order to show that the reduced systems for the Einstein--scalar field 
system and the Einstein--stiff fluid systems, are in Fuchsian form, 
we need to show that 
$t^{2-\alpha^a_{\ b}} M^a_{\ b} = \ordo(t^{\delta})$ for some $\delta > 0$. 

We consider first the Einstein--scalar field case. In this case, 
$$
M^a_{\ b} = g^{ac} e_c(\phi) e_b(\phi)
$$

Arguing as above for the estimate of $t^{2-\alpha^a_{\ b}}\SR^a_{\ b}$, 
we have 
$$
t^{2-\alpha^a_{\ b}} M^a_{\ b} \oleq 
 t^{2-\talpha^a_{\ b} - \eps} \tM_{ab}
$$
Therefore it is enough to show 
$$
t^{2-\talpha^a_{\ b}} \tM_{ab} \oleq t^{2\eps}
$$
By definition $\tM_{ab} = t^{-q_a - q_b} e_a(\phi)e_b(\phi)$ and hence 
$$
\tM_{ab} \oleq t^{-q_a - q_b}
$$
This give using $\talpha^a_{\ b} = |q_a - q_b|$, 
$$
t^{2 - \talpha^a_{\ b}} M^a_{\ b} \oleq t^{3q_1}
$$
For the scalar field case this gives, together with the above,  
$$
t^{2-\alpha^a_{\ b}} ( \SR^a_{\ b} - M^a_{\ b} ) \oleq t^{\delta}, 
\quad \text{ for some $\delta > 0$ }
$$
which is the estimate required for proving that the second reduced system for
the Einstein--scalar field system is in Fuchsian form. The argument for the
Einstein--stiff fluid system is very similar.  

\section{The constraints}\label{sec:constraints}

The main aim of this section is to show that if a solution of the evolution
equations is given which corresponds to a solution of the velocity dominated
equations as in Theorem \ref{maintheorem} or \ref{stiffmaintheorem} then 
it satisfies the full constraints. It will also be shown how the existence
of a large class of solutions of the velocity dominated constraints can be
demonstrated. 

The first result on the propagation of the constraints relies on rough
computations which prove the result in the case where all $p_a$ are close
to $1/3$.  An analytic continuation argument then gives the general case.

\begin{lemma}\label{smallconstraints}
Let $({}^0g_{ab},{}^0k_{ab},{}^0\phi)$ be a solution of the velocity dominated 
system as in the hypotheses of Theorem \ref{maintheorem} with 
$|p_a-1/3|<\alpha_0/10$.
Let $(\gamma^a{}_b,\kappa^a{}_b,\psi)$ be a solution of 
(\ref{eq:ein-first-red}) and (\ref{eq:wave-first}) modelled on this velocity 
dominated solution and define 
$g_{ab}$, $k_{ab}$ and $\phi$ by (\ref{ansatz-intro}). Suppose that this 
solution satisfies the properties 1.-6. of the conclusions of the Theorem
\ref{maintheorem}. Then the Einstein constraints are also satisfied.
\end{lemma}

\noindent
{\bf Proof} Define:
\begin{align}
C&=-k_{ab}k^{ab}+(\tr k)^2-R-16\pi\rho      \\
C_b&=\nabla_a k^a{}_b-\nabla_b(\tr k)-8\pi j_b
\end{align}
These quantities satisfy the evolution equations
\begin{align}
\partial_t C-2(\tr k)C&=\nabla^a C_a                               \\
\partial_t C_a-(\tr k)C_a&=\frac{1}{2}\nabla_a C
\end{align}
Define rescaled quantities by $\bar C=t^{2-\eta_1}C$ and $\bar C
=t^{1-\eta_2}C_a$ for some positive real numbers $\eta_1$ and $\eta_2$. Then 
the above equations can be written in the form:
\begin{align}
t\d_t\bar C+\eta_1\bar C&=2[1+t\tr k]\bar C-t^{2-\eta_1+\eta_2}
\nabla^a \bar C_a         \\
t\d_t\bar C_a+\eta_2\bar C_a&=[1+t\tr k]\bar C_a-(1/2)
t^{\eta_1-\eta_2}\nabla_a\bar C
\end{align}
Choose $\eta_1$ and $\eta_2$ so that $\eta_1-\eta_2>0$. The aim is to apply
Theorem \ref{fuchstheorem} to show that $\bar C$ and $\bar C_a$ vanish. In
order to do this we should show that these two quantities vanish as fast as
a positive power of $t$ as $t\to 0$, that $1+t\tr k$ vanishes like a positive
power of $t$ and that the term $\nabla^a \bar C_a$ is not too singular. In
obtaining these estimates it is necessary to use the behaviour of the 
derivatives of the solution mentioned in the remark following Theorem
\ref{fuchstheorem}. Note that since the velocity dominated constraints are 
satisfied by assumption, it is enough to estimate the differences of the 
constraint quantities corresponding to the velocity dominated and full 
solutions, since these are in fact equal to $C$ and $C_a$. By 
property 2. of the conclusions of Theorem \ref{maintheorem} it follows
that $1+t\tr k=O(t^{\alpha_0})$ which gives one of the desired statements.
Similarly it follows that 
\begin{equation}
-k^{ab}k_{ab}+(\tr k)^2=-({}^0k^{ab})({}^0k_{ab})+(\tr{}^0 k)^2
+O(t^{-2+\alpha_0})
\end{equation}
It follows from the curvature estimates done in section \ref{sec:curvest}
that the scalar
curvature is also $O(t^{-2+\alpha_0})$. Now consider the expression 
$\rho-{}^0\rho$. The components of the inverse metric can be estimated by
$t^{-2+40\alpha_0}$, so that the terms in $\rho$ involving spatial
derivatives can be estimated by $t^{-2+\alpha_0}$ as well. The 
difference of the time derivatives can be estimated by $t^{-2+\beta}$.
These are the estimates for the Hamiltonian constraint that will be needed. 

The estimates just carried out were independent of the restriction on the 
$p_a$ in the hypotheses of the lemma. The following estimates for the
momentum constraint are of a cruder type and do use the restriction.
First note that the metric and its inverse can be estimated
by the powers of $t$ equal to $2p_1$ and $-2p_3$ respectively. 
It follows from the assumption on the $p_a$ in the hypotheses of the 
theorem that $2p_1>2/3-\alpha_0/5$ and $-2p_3>-2/3-\alpha_0/5$. Using
(\ref{eq:conncoeff}) then shows that the connection coefficients can be 
estimated in terms of the power $-2\alpha_0/5$. The effect on the
order of a term of taking a divergence can be estimated by the powers
$-2\alpha_0/5$ and $-2/3-3\alpha_0/5$ for upper and lower indices 
respectively.

The gradient of the mean curvature is $O(t^{-1+\alpha_0}\log t)$ while the
difference of $j_a$ is $O(t^{-1+\beta}\log t)$. The difference of the 
divergence of the second fundamental form produces the power $-1+3\alpha_0/5$. 
Evidently the last power and that containing $\beta$ are the limiting ones 
and determine the estimate for $C_a$. Similarly the divergence of $C_a$ can 
be estimated by the powers $-5/3$ and $-5/3+\beta-3\alpha_0/5$. Note that it 
follows from the definition of $\alpha_0$ that $\alpha_0<1/30$. Thus given the 
hypothesis of the lemma it can be concluded that $\eta_1$ and $\eta_2$ can be 
chosen in such a way that all terms on the right hand side of the propagation 
equations for the constraint quantities vanish like positive powers of $t$. 
Thus these equations are Fuchsian and the conclusion follows from Theorem
\ref{fuchstheorem}. 

\vskip 10pt\noindent
The analogue of this lemma with the scalar field replaced by a stiff fluid
is also true and can be proved in the same way. Next the restriction on the
exponents $p_a$ will be lifted. Consider a solution $({}^0g_{ab},
{}^0k_{ab},{}^0\mu,{}^0v_a)$ of the velocity dominated constraints for the 
Einstein-stiff fluid system. Let ${}^0\hat k_{ab}$ be the trace-free part of
${}^0 k_{ab}$. The velocity dominated constraints become:
\begin{align}
-{}^0\hat k^{ab}{}^0 \hat k_{ab}+(2/3)(\tr {}^0 k)^2&=16\pi\mu        \\
\nabla_a {}^0\hat k^a_{\ b}&=8\pi\mu u_b                  
\end{align}
Now let ${}^\lambda k_{ab}=(1-\lambda)\hat {}^0k_{ab}+(1/3)
(\tr {}^0 k){}^0 g_{ab}$ and
\begin{align}
{}^\lambda\mu&=(1/16\pi)[-(1-\lambda)^2({}^0\hat k^{ab})({}^0\hat k_{ab})
+(2/3)(\tr {}^0k)^2]               \\
{}^{\lambda}\mu_b&=2(1-\lambda)\nabla^a{}^0\hat k_{ab}
[-(1-\lambda)^2({}^0\hat k^{ab})({}^0\hat k_{ab})+(2/3)(\tr {}^0k)^2]^{-1}
\end{align}
Then ${}^0 g_{ab}, {}^\lambda k_{ab}, {}^\lambda\mu, {}^\lambda u_a$ is a 
one parameter family of solutions of the velocity dominated constraints
which depends analytically on the parameter $\lambda$. There exists a
corresponding family of solutions of the velocity dominated evolution 
equations which also depends analytically on $\lambda$. Next, Theorem 
\ref{fuchstheorem} provides a corresponding analytic family of solutions of 
the full evolution equations. (Cf. the second remark following that theorem.) 
These define constraint quantities depending analytically on 
$\lambda$. For $\lambda$ close to one Lemma \ref{smallconstraints} shows
that these quantities are zero. Hence by analyticity they are zero for all
values of $\lambda$, including $\lambda=0$. This means that the conclusion
of Lemma \ref{smallconstraints} holds for all positive $p_a$ and a stiff 
fluid. Since any solution of the second reduced system for a scalar field
defines a solution of the second reduced system for a stiff fluid, this 
extension also holds for the scalar field.

A variant of the conformal method for solving the full Einstein constraints
can be used to analyse the velocity dominated constraints. Consider the
following set of free data: a Riemannian metric $\bar g_{ab}$, a symmetric 
trace-free tensor $\sigma_{ab}$ on $S$ and two scalar functions $\bar\phi$ 
and $\bar\phi_t$ on $S$. Next consider the following ansatz:
\begin{align}
g_{ab}&=\omega^4\bar g_{ab}                            \\
k_{ab}&=-(1/3)t_0^{-1}g_{ab}+\omega^{-2}l_{ab}                \\   
\phi&=\omega^{-2}\bar\phi                              \\
\phi_t&=\omega^{-4}\bar\phi_t
\end{align}
where
\begin{equation}
l_{ab}=\sigma_{ab}+\nabla_aW_b+\nabla_bW_a
-(2/3)\bar g_{ab}\bar g^{cd}\nabla_cW_d                         
\end{equation}
Putting this into (\ref{vdemrho}) and defining $\bar\rho=\frac{1}{2}
(\bar\phi_t)^2$ gives $\rho=\omega^{-8}\bar\rho$, a relation well known
from the usual conformal method. As a result of the Hamiltonian constraint
the function $\phi$ satisfies the following algebraic analogue of the 
Lichnerowicz equation:
\begin{equation}
-\omega^{-12}l_{ab}l_{cd}\bar g^{ac}\bar g^{bd}+\frac{2}{3}t_0^{-2}
-16\pi\omega^{-8}\bar\rho=0
\end{equation}
Solving this comes down to looking for positive roots of the equation
$a\zeta^3+b\zeta^2-c=0$ where $a$ and $b$ are non-negative and $c$ is
positive. The derivative of the function on the left hand side of this 
equation is $\zeta(3a\zeta+2b)$. Thus unless $a$ and $b$ are both zero
the derivative has no positive roots. Moreover the function tends to 
plus infinity for large $\zeta$ and is negative at $\zeta=0$. 
Hence the equation has a unique solution for each $a$ and $b$ not both
zero and if $a$ and $b$ depend analytically on some parameter then the 
solution does so too. If $a$ and $b$ are both zero then of course there 
is no positive solution. In the case of interest here $a$ and $b$ are both 
positive. The function $\omega$ is given by $\omega=\Omega(l_{ab}
l_{cd}\bar g^{ac}\bar g^{bd},t_0,\bar\rho)$ where the analytic 
function $\Omega$ is defined as the solution of the algebraic Lichnerowicz 
equation. The momentum constraint implies the elliptic equation
\begin{equation}\label{york}
\bar g^{as}\nabla_a[\nabla_s W_b+\nabla_b W_s-(2/3)\bar g_{sb}
\bar g^{cd}\nabla_c W_d]=8\pi[\bar j_b-2\bar\phi_t\bar\phi
\omega\nabla_b\omega]
-\nabla_a\sigma^a{}_b
\end{equation}
for $W_a$. Here $\bar j_a=\bar\phi_t\nabla_a\bar\phi$. Note that,
when $\omega$ is expressed in terms of the function $\Omega$ of the basic
variables, it depends on the first derivatives of $W_a$. Thus the 
expression $\nabla_a\omega$ involves second derivatives of $W_a$ and is
not simply a lower order term.

Consider now the linearization of (\ref{york}),with respect to $W_a$,
where $\omega$ has been reexpressed using $\Omega$. In particular,
consider the linearization in the particular case where $\bar g_{ab}$ is 
the metric of constant negative curvature on a compact hyperbolic manifold, 
the tensor $\sigma_{ab}$ is zero, $\bar\phi$ and $\bar\phi_t$ are
constant and the background value of $W_a$ is zero. Because $\omega$ 
is a function of an expression quadratic in $W_a$, the right hand side
of (\ref{york}) makes no contribution to the linearization. Since 
$\bar g_{ab}$ has no non-trivial conformal Killing vectors it follows
from the standard theory of the York operator that the operator obtained
by linearization of the equation (\ref{york}) is invertible as a map
between appropriate Sobolev spaces. Then an application of the
implicit function theorem gives solutions of (\ref{york}) for 
arbitrary choices of the free data sufficiently close (with respect to
a Sobolev norm) to the particular free data at which the linearization was 
carried out. This shows the existence of solutions of the velocity dominated
constraints which are as general as the solutions of the full Einstein
constraints (at least in the crude sense of function counting).

The conformal method can be applied in a similar way in the stiff fluid case 
and it turns out 
to be easier than in the scalar field case. This might seem paradoxical,
since the scalar field problem can be identified with a subcase of the stiff
fluid problem. The explanation is that it is difficult to identify which
free data in the procedure for constructing stiff fluid data which will
be presented correspond to data for a scalar field. The ansatz used 
is $\mu=\omega^{-8}\bar\mu$ and $u_a=\omega^2\bar u_a$. This gives
the scaling $\rho=\omega^{-8}\bar\omega$ and $j_a=\omega^{-6}\bar j_a$
which is often used in the conformal method. The quantities describing the
geometry are scaled as in the case of the scalar field. The equations for
$\omega$ and $W_a$ are very similar in both cases, with the notable 
difference that in the stiff fluid case the term involving the derivative of
$\omega$ is missing from the equation for $W_a$. This means that the 
equation for $W_a$ is independent of $\omega$ and can be solved by standard
theory, as long as the metric $\bar g_{ab}$ has no conformal Killing
vectors. Once this has been done the algebraic equation for $\omega$ can
be solved straightforwardly.

\section{Discussion}\label{sec:discussion}

We have shown the existence of a family of solutions of the Einstein 
equations coupled to a scalar field or a stiff fluid whose singularity
structure we can analyse. No symmetry assumptions are made and the solutions
are general in the sense that they depend on the same number of free 
functions as general initial data for the same system on a regular Cauchy 
surface. These solutions agree with the picture of general spacetime 
singularities proposed by Belinskii, Khalatnikov and Lifshitz in two 
important ways. Firstly, the evolution at different spatial points decouples, 
in the sense that the solutions of the full equations are approximated near 
the singularity by a solution of a system of ordinary differential equations.
Secondly there exists a Gaussian coordinate system which covers a 
neighbourhood of the singularity in which the singularity is situated at
$t=0$. It is easily seen that the curvature invariant $R_{\alpha\beta}
R^{\alpha\beta}=64\pi^2(\nabla_\alpha\phi\nabla^\alpha\phi)^2$ blows up 
uniformly for $t\to 0$. In fact the leading term is proportional to
$A^4(x)t^{-4}$ and in the solutions
we consider $A$ can never vanish, as a consequence of the Hamiltonian
constraint. Thus these singularities are all consistent with the strong
cosmic censorship hypothesis. The mean curvature of the hypersurfaces
of constant Gaussian time tends uniformly to infinity as $t\to 0$ so
that the the singularity in crushing in the sense of \cite{eardley79}.
It then follows from well-known results that a neighbourhood of the
singularity can be covered by a foliation consisting of constant mean 
curvature hypersurfaces. This is the most general class of spacetimes
in which all these suggested properties of general spacetimes have been
demonstrated. A subclass of these spacetimes is covered by the results of
Anguige and Tod\cite{anguige99}. The connection between their results and
those of the present paper deserves to be examined more closely but
intuitively their spacetimes should correspond to the case where, in our
notation, the $p_i$ are everywhere equal to $1/3$. 

The spacetimes constructed have been shown to be general in the sense of
function counting. It would, however, be desirable to prove that the
assumption of analyticity of the data can be replaced by smoothness and
that, this having been done, the spacetimes constructed include all those
arising from a non-empty open set of initial data on a regular Cauchy
surface which, in particular, contains the initial data for a Friedmann
model. This would be a statement on the stability of the Friedmann
singularity. A model for this kind of generalization is provided by
the work of Kichenassamy\cite{kichenassamy96} on nonlinear wave equations.

It was indicated in the introduction that the results on the Einstein-scalar
equations can be interpreted in more than one way. The interpretation
which has been emphasized here is that where the metric occurring in this
system is considered to be the physical metric.  In the interpretation in
terms of string cosmology the physical metric is (up to a multiplicative
constant) $e^\phi g_{\mu\nu}$. This means that for $A(x)$ sufficiently
negative the limit $t\to 0$ does not correspond to a singularity at all,
but rather to a phase which lasts for an infinite proper time. It is the
time reverse of this situation which plays a role in the pre-big bang
model\cite{buonanno99}. Another interpretation is in terms of the vacuum
field equations in Brans-Dicke theory. This is very similar to the string
cosmology case, with the difference that the conformal factor $e^\phi$ is 
replaced by $e^{C\phi}$ where $C$ is a constant which depends on the 
Brans-Dicke coupling constant.

All the results in this paper have concerned the case of three space 
dimensions. There are reasons to believe that if the space dimension is
at least ten then the vacuum Einstein equations allow stable quiescent 
singularities, similar in some ways to those of the Einstein-scalar field
equations in three space dimensions\cite{demaret85}. The techniques developed 
in this paper might allow this to be proved rigorously. It would also to
be interesting to know what happens to the picture when further matter
fields are added. There are several possibilities here. One is to add some
other field, not directly coupled to the scalar field, to the 
Einstein-scalar field system. A second is to reinstate some of the
extra fields (axion, moduli) which have been discarded in
passing from the low energy limit of string theory to the Einstein-dilaton
theory. A third is to add extra matter fields to the Brans-Dicke theory.

Another direction in which the results on the Einstein-scalar field and
Einstein-stiff fluid equations could be generalized is to start with 
situations where the solution has one Kasner exponent negative
and investigate whether it moves (in the direction towards the
singularity) towards the region where all Kasner exponents are 
non-negative. If this were true, then the singularities in generic
solutions of these equations could be quiescent. The set of initial data
concerned would be not just open, but also dense. This question is
sufficiently difficult that it would seem advisable to first try and
investigate it rigorously in the spatially homogeneous case.

\vskip 10pt\noindent
{\it Acknowledgements} This research was supported in part by the Swedish
Natural Sciences Research Council (SNSRC), contract no. F-FU 4873-307, and
the US National Science Foundation under Grant No. PHY94-07194. Part of 
the work was done while the authors were enjoying the hospitality of the 
Institute for Theoretical Physics, Santa Barbara. We gratefully acknowledge 
stimulating discussions with V. Moncrief which had an important influence on 
the development of the strategy used in this work.

\end{document}